\documentclass[aps,prl,superscriptaddress,amsmath,amssymb,floatfix,reprint,twocolumn]{revtex4-2}
\pdfoutput=1

\usepackage{amssymb}
\usepackage{graphicx}
\usepackage{bm}
\usepackage[pdftex,breaklinks=true,bookmarksopen=true,bookmarksopenlevel=3,bookmarksnumbered=true,colorlinks=true,urlcolor= magenta,citecolor=blue,linkcolor=blue]{hyperref}
\usepackage{float}
\usepackage[utf8]{inputenc}
\usepackage[T1]{fontenc}
\usepackage{verbatim}
\usepackage{siunitx}
\usepackage[dvipsnames]{xcolor}
\usepackage{soul}
\usepackage{ulem}
\usepackage{slashed}
\usepackage{lineno}

\begin{document}

\title{Self-triggered strong-field QED collisions in laser-plasma interaction}

\author{Aimé Matheron}
\affiliation{LOA, ENSTA Paris, CNRS, Ecole Polytechnique, Institut Polytechnique de Paris, 91762 Palaiseau, France}
\author{Igor Andriyash}
\affiliation{LOA, ENSTA Paris, CNRS, Ecole Polytechnique, Institut Polytechnique de Paris, 91762 Palaiseau, France}
\author{Xavier Davoine}
\affiliation{CEA, DAM, DIF, 91297 Arpajon, France}
\affiliation{Universit\'{e} Paris-Saclay, CEA, LMCE, 91680 Bruy\`{e}res-le-Ch\^{a}tel, France}
\author{Laurent Gremillet}
\affiliation{CEA, DAM, DIF, 91297 Arpajon, France}
\affiliation{Universit\'{e} Paris-Saclay, CEA, LMCE, 91680 Bruy\`{e}res-le-Ch\^{a}tel, France}
\author{Mattys Pouyez}
\affiliation{LULI, Sorbonne Université, CNRS, CEA, Ecole Polytechnique, Institut Polytechnique de Paris, F-75255 Paris, France}
\author{Mickael Grech}
\affiliation{LULI, CNRS, CEA, Sorbonne Université, Ecole Polytechnique, Institut Polytechnique de Paris, F-91120 Palaiseau, France}
\author{Livia Lancia}
\affiliation{LULI, CNRS, CEA, Sorbonne Université, Ecole Polytechnique, Institut Polytechnique de Paris, F-91120 Palaiseau, France}
\author{Kim Ta Phuoc}
\affiliation{Univ. Bordeaux, CNRS, CEA, CELIA (Centre Lasers Intenses et Applications), UMR 5107, F-33400 Talence, France}
\author{Sébastien Corde}
\email[Corresponding authors:\\ ]{aime.matheron@polytechnique.edu\\ sebastien.corde@polytechnique.edu}
\affiliation{LOA, ENSTA Paris, CNRS, Ecole Polytechnique, Institut Polytechnique de Paris, 91762 Palaiseau, France}
\affiliation{SLAC National Accelerator Laboratory, Menlo Park, CA 94025, USA}

\begin{abstract}

Exploring quantum electrodynamics in the most extreme conditions, where electron-positron pairs can emerge in the presence of a strong background field, is now becoming possible in Compton collisions between ultraintense lasers and energetic electrons. In the strong-field regime, the colliding electron emits $\gamma$ rays that decay into pairs in the strong laser field. While the combination of conventional accelerators and lasers of sufficient power poses significant challenges, laser-plasma accelerators offer a promising alternative for producing the required multi-GeV electron beams. To overcome the complexities of colliding these beams with another ultraintense laser pulse, we propose a novel scheme in which a single laser pulse both accelerates the electrons and collides with them after self-focusing in a dedicated plasma section and reflecting off a plasma mirror. The laser intensity boost in the plasma allows the quantum interaction parameter to be greatly increased. Using full-scale numerical simulations, we demonstrate that a single \SI{100}{J} laser pulse can achieve a deep quantum regime with electric fields in the electron rest frame as high as $\chi_e\sim 5$ times the Schwinger critical field, resulting in the production of about 40 pC of positrons.
\end{abstract}

\maketitle

\noindent
Recent progress in laser technology \cite{Danson_HPLSE_2015, Danson2019, Yoon_Optica_2021, Radier_HPLSE_2022} have revived strong interest in experimental probing of strong-field quantum electrodynamics (SFQED)~\cite{Nikishov_JETP_1964, Erber_RMP_1966, DiPiazza_RMP_2012} in laser-electron interactions \cite{Bell_PRL_2008, Ridgers_PRL_2012, Vranic_PRL_2014, Lobet_PRAB_2017, Magnusson_PRL_2019,  Vincenti_PRL_2019, Lecz_PPCF_2019, Blackburn_RMPP_2020, Mercuri_Baron_2021, Qu_PRL_2021, PhysRevLett.124.044801,BaumannQED}.
In this regime, ultrarelativistic electrons colliding with an intense enough laser experience a boosted electromagnetic (EM) field with $\chi_e\gtrsim 1$, where $\chi_e =E^\star/E_\mathrm{S}$ is the electron quantum parameter that measures how deep the interaction is in the quantum regime. Here, $E^\star$ denotes the electric field in the electron rest frame and $E_\mathrm{S}=m_e^2c^3/(q_e \hbar)\simeq \SI{1.3e18}{V.m^{-1}}$ is the Schwinger critical field \cite{Schwinger_PR_1951} ($m_e$ and $q_e$ are the electron mass and charge, $\hbar$ is the reduced Planck constant, and $c$  the speed of light). 
When $\chi_e$ approaches 1 and the electromagnetic fields at play become very high, the usual perturbative approach of regular QED breaks down because of the no-longer negligible high order interaction terms between the Dirac field and the EM field. 
The theoretical framework of SFQED, characterized by $\chi_e \gtrsim 1$, relies on a classical description of the laser field in the so-called Furry picture and redefines a new spinor basis in which the in and out vacuum spinor states are replaced by Volkov's dressed states~\cite{Volkov1935,Berestetskii1982}, corresponding to particle states in the strong field. Feynman rules for SFQED can be computed from these dressed states, giving rise to new or modified processes, in particular nonlinear inverse Compton scattering (NICS) in a strong field where an electron absorbs $n$ laser photons and emits a high-energy $\gamma$ photon ($e^-+ n\omega_{\mathrm{las}} \rightarrow e^- + \gamma$), and nonlinear Breit-Wheeler (NLBW) where the $\gamma$ photon and $n$ laser photons convert into an electron-positron pair via a pure light-by-light interaction $\gamma+n\omega_{\mathrm{las}} \rightarrow e^-+e^+$ \cite{Kirk_PPCF_2009, DiPiazza_RMP_2012}.

The theoretical development of SFQED received major contributions a few decades ago from Ritus, Narozhny and others \cite{Ritus_AP_1972, Narozhny_PRD_1980, Fedotov_JPhCS_2017}. 
However, experimental measurements of SFQED have been restrained due to the extreme laser intensities required.
The experimental difficulty of breaking the quantum vacuum to create pairs lies in the exponential suppression of the NLBW cross section as $\chi_\gamma \exp\left(-8/3\chi_\gamma\right)$ at $\chi_\gamma\ll1$~\cite{Fedotov2023}, that is, at low photon energy and field strength. 
Here $\chi_\gamma=(2\hbar\omega_\gamma/m_ec^2)(E/E_{\mathrm{S}})$ is the photon quantum parameter, with $\hbar\omega_\gamma$ the $\gamma$ photon energy and $E$ the electric field in the laboratory frame seen by the $\gamma$ photon. 
To date, the only experimental measurement of positrons created via multiphoton Breit-Wheeler was performed at SLAC \cite{Burke_PRL_1997, Bamber_PRD_1999} where a 47 GeV electron beam collided with a $\SI{5e17}{W.cm^{-2}}$ laser pulse. The electron quantum parameter reached up to $\sim 0.3$ with a total of hundred positrons detected throughout the duration of this seminal experiment and a production rate reaching up to 0.2 positron per shot. 

With the development of multi-petawatt, ultraintense laser systems, it is now envisioned to probe electron-laser collisions at $\chi_e > 1$ in all-optical experiments based on laser-plasma accelerators (LPAs)~\cite{Tajima_PRL_1979, Faure_Nature_2004, Geddes_Nature_2004, Mangles_Nature_2004}. The latter use an ultraintense laser pulse to drive a charge-density wave (the so-called plasma wave) that can sustain accelerating electric fields orders magnitude higher than those achievable with conventional (radiofrequency) accelerators. Experimentally, LPAs are now able to deliver electron energies up to 10 GeV~\cite{PhysRevLett.122.084801, 10.1063/5.0161687}. The standard approach for LPA-based SFQED experiments consists of accelerating electrons with a first laser pulse (the LPA driver) and then colliding them with a second intense laser pulse to trigger SFQED processes~\cite{Poder_PRX_2018, Cole_PRX_2018, Blackburn_RMPP_2020}. However, this scheme is hindered by shot-to-shot fluctuations and very demanding alignment and synchronization requirements between the electron beam and the second laser pulse, which have thus far prevented substantial experimental progress.
\begin{figure}[b!]
    \centering
    \includegraphics[scale=0.15]{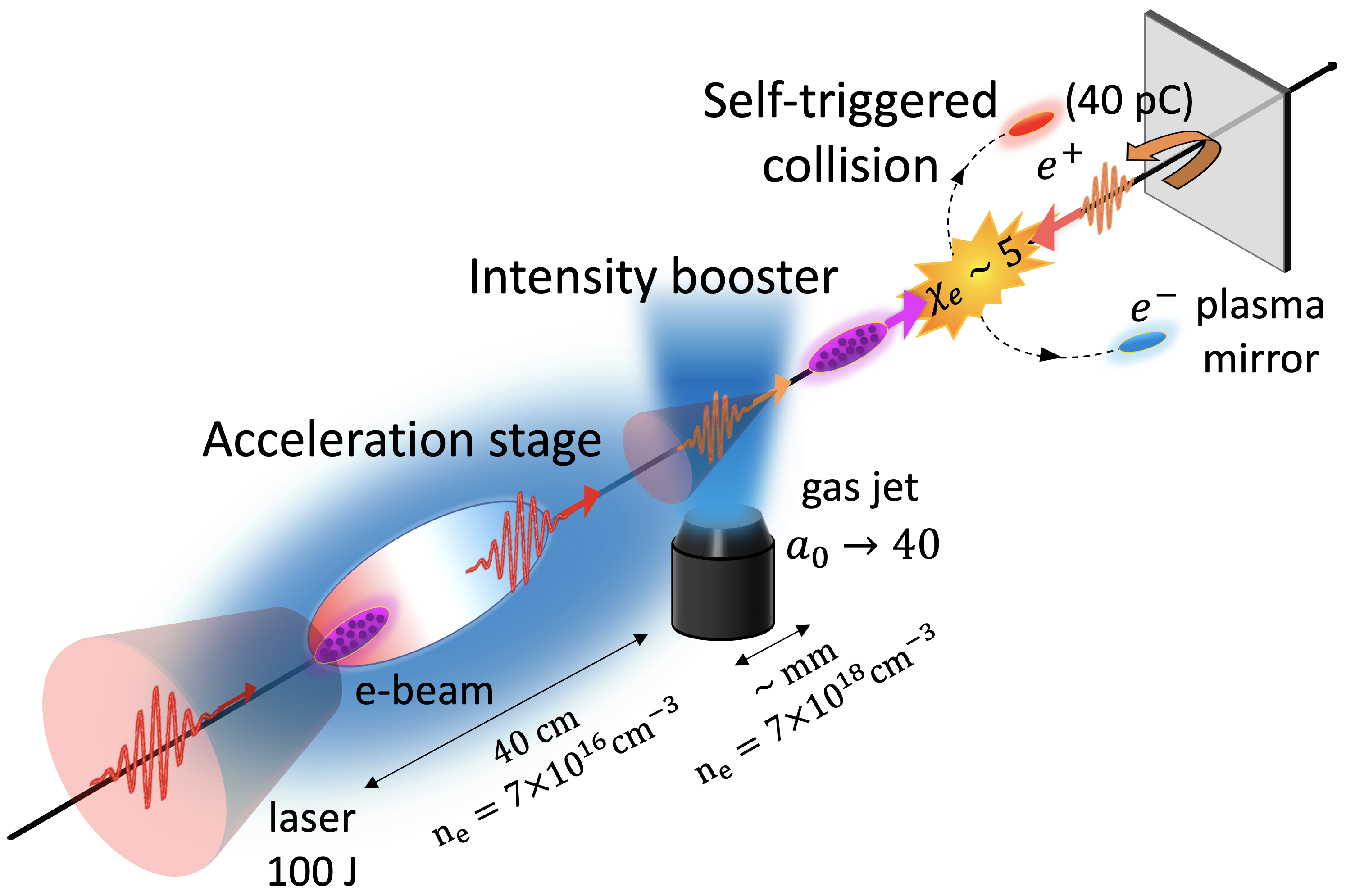}
    \caption{Concept for self-triggered Compton collisions in the SFQED regime. A multi-PW laser pulse first drives a laser-plasma accelerator to produce a high-energy electron beam in a low-density underdense plasma (acceleration stage). It is then intensified via self-focusing in a high-density underdense plasma (intensity booster). Finally, it is reflected by an overdense plasma (a foil acting as a plasma mirror) and collides back with the electron beam. This collision takes place in the SFQED regime, leading to massive $\gamma$-ray photon and pair generation through NICS and NLBW.}
    \label{fig1}
\end{figure}

Here we present a novel concept that considerably simplifies the realization of LPA-based SFQED experiments with two critical advantages that should allow SFQED to be investigated in a much more systematic way than currently envisaged: first, by making use of a single laser beam only, and second, by ensuring automatic spatial and temporal overlap between the colliding electron and laser beams. This is done by placing a plasma mirror (a solid foil ionized by the laser) in the path of the laser pulse exiting the LPA in order to reflect it back into the trailing electron beam propagating just behind it~\cite{TaPhuoc2012}. However, the ultrarelativistic laser pulse required to induce SFQED processes in the Compton collision is not optimal for the LPA, which usually works best at mildly relativistic intensities. To circumvent this problem, we add a high-density underdense plasma just after the LPA to enhance the laser intensity through relativistic self-focusing of the laser pulse. This intensity booster allows to reach laser intensities nearly two orders of magnitude higher than the initial vacuum laser intensity, so that the perfectly aligned Compton collision triggered by the plasma mirror can enter the SFQED regime when using a multi-PW laser. 

The proposed scheme, depicted in Fig.~\ref{fig1}, therefore  comprises three sections: (i) a low-density underdense plasma to generate an electron beam at the 10-GeV scale, (ii) a high-density underdense plasma to boost the intensity, and (iii) an overdense plasma mirror to bring about the Compton interaction with \SI{100}{\%} collision probability. We emphasize that the intensity booster is pivotal to this setup, amplifying the effective $\chi_e$ by nearly an order of magnitude and thus enabling the Compton collision to occur under deep SFQED conditions. 

\begin{figure*}[t!]
    \centering
    \includegraphics[width=17cm]{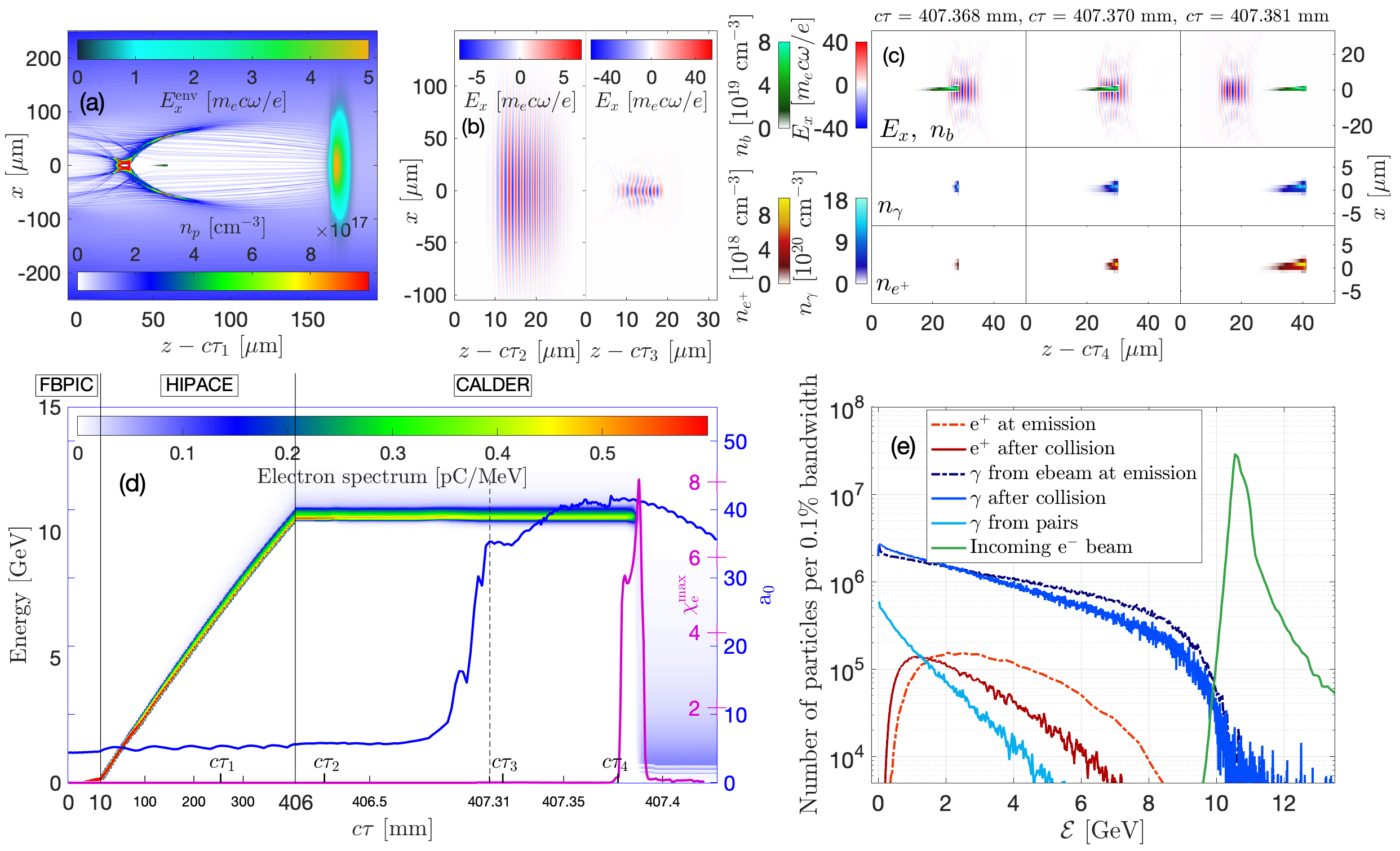}
    \caption{Start-to-end, full-scale numerical model of the single-laser SFQED setup. (a) Spatial distributions of the laser pulse (characterized by its normalized electric field envelope), electron plasma density and electron beam in the LPA stage (HiPACE++ output at $c\tau_1=\SI{250}{mm}$). (b) Spatial distribution of the laser field $E_x$ before (left) and after (right) self-focusing (CALDER output at $c\tau_2=\SI{406.2}{mm}$ and $c\tau_3=\SI{407.32}{mm}$). (c) Successive snapshots of the Compton collision, showing the laser electric field $E_x$ and electron beam density $n_b$ (first row), the positron density $n_{e^+}$ (second row) and the $\gamma$-ray photon density $n_\gamma$ (third row) (CALDER outputs at $c\tau=\SI{407.368}{mm}$, \SI{407.370}{mm}, \SI{407.381}{mm}). (d) Time evolution of the electron beam energy spectrum, laser $a_0$ and maximum of $\chi_e$ during the three interaction stages. (e) Energy spectra of the $\gamma$-ray photons and positrons at their times of emission and after the Compton collision, as well as incoming electron spectrum.}
    \label{Fig2}
\end{figure*}

To demonstrate its potential, we have developed a complete start-to-end numerical model of this scheme, by combining three particle-in-cell simulation codes to describe the relevant kinetic and SFQED processes for each interaction stage (see Supplemental Material for details on the simulation codes and parameters). The results are summarized in Fig.~\ref{Fig2}. In detail, we have considered a Gaussian laser pulse with \SI{0.8}{\mu m} central wavelength, \SI{100}{J} energy, and \SI{20}{fs} full-width-at-half-maximum (FWHM) duration, focused to a \SI{100}{\mu m} (FWHM) spot at the entrance of the low-density underdense plasma (LPA stage). Its peak intensity is \SI{4.1e19}{W.cm^{-2}}, corresponding to a normalized vector potential $a_0=4.4$. A plasma channel is used to guide the laser pulse, with a radial density profile of the form $n_e(r)=n_e^0\left(1+ \frac{r^2}{r_c^2}\right)$ with $r_c=\SI{182}{\mu m}$ the characteristic channel radius and $n_e^0$ the on-axis plasma density. The LPA is simulated in two steps. The first models the injection of the electron beam by means of the FBPIC code~\cite{Lehe2016}, using a plasma density downramp, with $n_e^0$ dropping linearly from \SI{1e17}{cm^{-3}} to \SI{7e16}{cm^{-3}} over a distance of \SI{200}{\mu\mathrm{m}}. This results in the injection of a $\SI{186}{pC}$ electron bunch in the simulation. After a propagation distance of \SI{1}{cm}, the interaction is described over an additional 40 cm with the HiPACE++ code~\cite{Diederichs2022}. The quasistatic approximation underlying this code allows the LPA to be treated at a very low computational cost. The on-axis plasma density is kept constant at $n_e^0=\SI{7e16}{cm^{-3}}$ in this second stage. We use the LASY tool~\cite{10.5281/zenodo.10091858} to transfer the beam electron and laser field data from FBPIC to HiPACE++.

Figure~\ref{Fig2}(a) illustrates the LPA stage with a snapshot of the laser pulse, electron plasma density and injected electron beam from HiPACE++.
Figure~\ref{Fig2}(d) plots the evolution of the electron beam spectrum and of the peak laser field strength $a_0$ as a function of the propagation distance. In the LPA stage, the electron beam energy increases linearly, reaching a final mean energy of 11 GeV, while the laser field strength does not vary substantially.

The central part of the concept, the intensity booster and SFQED collision, is then modeled with the three-dimensional (3D) PIC-QED code CALDER~\cite{Lefebvre_NF_2003,Lobet_JPCS_2016} using the electron beam and laser fields from HiPACE++ as an input. Here the laser intensity envelope and the linear frequency chirp induced in the LPA are accounted for in the code transition. A single CALDER simulation (Fig.~\ref{Fig2} for $c\tau \ge \SI{406}{mm}$) is performed to describe both the nonlinear dynamics of the laser and electron beams in the high-density underdense plasma and their collision following the reflection of the laser off the plasma mirror. A moving simulation window is used throughout the underdense plasma. The density profile of the latter first rises linearly from the LPA density (\SI{7e16}{cm^{-3}}) to a value of \SI{7e18}{cm^{-3}} over \SI{250}{\mu m}, and plateaus afterwards over a distance of \SI{1.5}{mm}. The plasma density is uniform transversely. Due to relativistic self-focusing, the laser shrinks to a spot size of approximately \SI{8}{\mu m} (FWHM) while its normalized field strength $a_0$ increases from 5.6 to 41 [see Fig.~\ref{Fig2}(b) for snapshots of the laser pulse before and after self-focusing, and Fig.~\ref{Fig2}(d) for the time evolution of $a_0$]. This corresponds to a $50\times$ intensity increase in the booster, and to a final intensity nearly two orders of magnitude larger than the initial vacuum laser intensity. The laser pulse duration at the exit of the underdense plasma is \SI{19}{fs} (FWHM), similar to its initial value. The electron beam remains approximately unperturbed by the high density over the small distance before its collision, mainly because of its initial high energy.

\begin{figure}[t!]
    \centering
    \includegraphics[width=8.5cm]{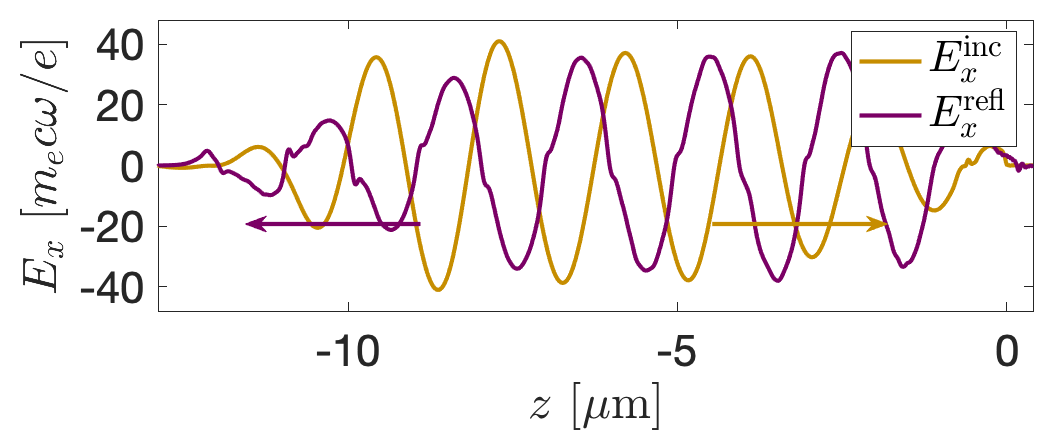}
\caption{Laser reflection on the plasma mirror. On-axis laser electric field for incident (yellow) and reflected (purple) pulses.}
    \label{fig3bis}
\end{figure}

The collision between the laser and electron  beams is triggered by applying a perfectly reflective boundary condition for the electromagnetic fields at the desired position ($z=\SI{407.45}{mm}$) of the plasma mirror. This method allows us to use the same spatial resolution as through the underdense plasma, thus keeping the computational cost reasonable. The simulation window no longer moves at this stage. The validity of the specular reflection approximation was verified in a separate simulation modeling only the laser-foil interaction with a mesh size resolving the skin depth at solid density (see Fig.~\ref{fig3bis} and Supplemental Material). The SFQED processes arising during the collision are modeled accurately using the SFQEDtoolkit~\cite{Montefiori2023} implemented into CALDER. Figure~\ref{Fig2}(d) reveals that the maximum of the electron quantum parameter first reaches a value of $\chi_e \simeq 5.5$, in good agreement with the theoretical value of $\chi_e=5.4$ expected for $\mathcal{E}=\SI{11}{GeV}$ and $a_0=41$, and later peaks to a value of 8 due to the presence of a small number of higher-energy electrons at the rear of the bunch [see high-energy tail in electron spectrum of Fig.~\ref{Fig2}(e)] which have negligible contribution to the photon and pair generation. During the collision, the energy spectrum of the beam electrons drops suddenly from \SI{11}{GeV} down to a few GeV due to strong $\gamma$-ray emission.

The head-on laser-electron collision is illustrated in Fig.~\ref{Fig2}(c) at three successive times, showing in particular the generation of $\gamma$ photons and positrons via NICS and NLBW, which copropagate with the incident electron beam. Besides the automatic alignment between the laser and electron beams that guarantees stable collisions, another key advantage of our concept can be seen in Fig.~\ref{Fig2}(c), namely that the transverse size ($\sim \SI{8}{\mu m}$ FWHM) of the self-focused laser is substantially larger than that of the electron beam ($\sim \SI{1.5}{\mu m}$ FWHM). As a result, all beam electrons experience the same strong laser fields and thus the same SFQED conditions. This is particularly important to avoid background contributions such as linear inverse Compton scattering from electrons outside the central laser spot, which can dominate $\gamma$-ray generation in other collision geometries~\cite{claveria2023commissioningmeasurementsinitialxray}. In turn, the emitted $\gamma$ rays undergo the same strong laser fields, and thus pair creation via NLBW is very efficient: the large amount of positrons produced, \SI{41}{pC}, i.e. \SI{22}{\%} of the initial \SI{186}{pC} charge of the electron beam, should be easily detectable in a proof-of-principle experiment. This performance is in excellent agreement with the analytical modeling of pair production in electron-laser collisions described in Ref.~\cite{Pouyez2024}, which predicts a charge ratio of 0.19 for $a_0=40$, $\mathcal{E}_{e^-}=\SI{11}{GeV}$, and a laser pulse duration of \SI{20}{fs} (FWHM).

The energy spectra of the $\gamma$ photons and positrons are shown in Fig.~\ref{Fig2}(e). During the collision, $\gamma$ photons are emitted with energies as high as that of the beam electrons ($\sim\SI{11}{GeV}$), as represented by the dashed dark blue curve. However, upon interacting with the strong laser field, those photons can rapidly decay into $e^- e^+$ pairs. As the probability of the NLBW process increases with $\chi_\gamma \propto \hbar \omega_\gamma$, one expects the high-energy part of the final photon spectrum to significantly differ from its shape at the time of photon emission, as is confirmed in Fig.~\ref{Fig2}(e). Here, the pair kinetic energy at emission, $\mathcal{E}^k_{\mathrm{pair}}=\hbar(\omega_\gamma+n\omega_{\mathrm{las}})-2m_ec^2$ is essentially equal to that of the decaying photons ($\hbar \omega_\gamma$) since the pair rest energy ($2m_e c^2$) and the extra energy from the laser photons ($n\hbar \omega_\mathrm{las}$) are negligible compared to the GeV-range photon energies. At their creation times, the positrons are characterized by a broad energy spectrum, peaked around \SI{2.4}{GeV} and extending up to \SI{8}{GeV} approximately [see dashed light red curve in Fig.~\ref{Fig2}(e)]. These energetic positrons also experience the intense laser field and in turn radiate part of their energy into $\gamma$ photons, as represented by the light blue curve. This extra radiation from pairs adds up to that from the primary beam electrons, preferentially in the low-energy part of the photon spectrum as seen in Fig.~\ref{Fig2}(e).  These $\gamma$ photons from primary pairs however do not contribute to the generation of a substantial number of secondary pairs. As a result of these radiation losses for pairs, the positron spectrum is shifted to lower energies, with a peak moving from \SI{2.4}{GeV} (at emission times) to \SI{1.2}{GeV} (after the collision).

Finally, the conclusions of this 3D PIC-QED CALDER simulation with perfect reflection have been validated in a separate simulation modeling the laser reflection on the plasma mirror and resolving the plasma skin depth at solid density (see details in Supplemental Material). This refined simulation predicts only a small degradation of the laser field sinusoidal profile and a slight (by \SI{7}{\%}) reduction in its peak amplitude, as shown in Fig.~\ref{fig3bis}, thus reducing $\chi_e$ by the same amount but still triggering an SFQED collision with $\chi_e \sim 5$.

In summary, we demonstrated a novel all-optical concept to achieve SFQED interactions with a single multi-PW laser in a considerably simplified setup. Free from alignment constraints and taking advantage of a laser intensity boost, it can reach a deep quantum regime at $\chi_e\sim 5$ with about \SI{40}{pC} of $e^+$ created with clean QED observables as all high-energy electrons and emitted $\gamma$ photons experience similar SFQED conditions.
The relative simplicity of using a single laser and the guaranteed alignment and synchronization with the plasma mirror make this scheme a very promising shortcut towards measurements of SFQED processes at $\chi_e >1$.

The work was supported by the ANR (g4QED project, Grant No. ANR-23-CE30-0011). We acknowledge the Grand Équipement National de Calcul Intensif (GENCI) and Très Grand Centre de Calcul (TGCC) for granting us access to the supercomputer Joliot-Curie on the Irene Rome and Skylake partitions, under Grants No. 2023-S1405spe00030, No. 2023-A0140510786, No. 2023-A0150510062, No. 2024-A0160510786 to run PIC simulations.


\begin{thebibliography}{46}%
\makeatletter
\providecommand \@ifxundefined [1]{%
 \@ifx{#1\undefined}
}%
\providecommand \@ifnum [1]{%
 \ifnum #1\expandafter \@firstoftwo
 \else \expandafter \@secondoftwo
 \fi
}%
\providecommand \@ifx [1]{%
 \ifx #1\expandafter \@firstoftwo
 \else \expandafter \@secondoftwo
 \fi
}%
\providecommand \natexlab [1]{#1}%
\providecommand \enquote  [1]{``#1''}%
\providecommand \bibnamefont  [1]{#1}%
\providecommand \bibfnamefont [1]{#1}%
\providecommand \citenamefont [1]{#1}%
\providecommand \href@noop [0]{\@secondoftwo}%
\providecommand \href [0]{\begingroup \@sanitize@url \@href}%
\providecommand \@href[1]{\@@startlink{#1}\@@href}%
\providecommand \@@href[1]{\endgroup#1\@@endlink}%
\providecommand \@sanitize@url [0]{\catcode `\\12\catcode `\$12\catcode
  `\&12\catcode `\#12\catcode `\^12\catcode `\_12\catcode `\%12\relax}%
\providecommand \@@startlink[1]{}%
\providecommand \@@endlink[0]{}%
\providecommand \url  [0]{\begingroup\@sanitize@url \@url }%
\providecommand \@url [1]{\endgroup\@href {#1}{\urlprefix }}%
\providecommand \urlprefix  [0]{URL }%
\providecommand \Eprint [0]{\href }%
\providecommand \doibase [0]{https://doi.org/}%
\providecommand \selectlanguage [0]{\@gobble}%
\providecommand \bibinfo  [0]{\@secondoftwo}%
\providecommand \bibfield  [0]{\@secondoftwo}%
\providecommand \translation [1]{[#1]}%
\providecommand \BibitemOpen [0]{}%
\providecommand \bibitemStop [0]{}%
\providecommand \bibitemNoStop [0]{.\EOS\space}%
\providecommand \EOS [0]{\spacefactor3000\relax}%
\providecommand \BibitemShut  [1]{\csname bibitem#1\endcsname}%
\let\auto@bib@innerbib\@empty
\bibitem [{\citenamefont {Danson}\ \emph {et~al.}(2015)\citenamefont {Danson},
  \citenamefont {Hillier}, \citenamefont {Hopps},\ and\ \citenamefont
  {Neely}}]{Danson_HPLSE_2015}%
  \BibitemOpen
  \bibfield  {author} {\bibinfo {author} {\bibfnamefont {C.}~\bibnamefont
  {Danson}}, \bibinfo {author} {\bibfnamefont {D.}~\bibnamefont {Hillier}},
  \bibinfo {author} {\bibfnamefont {N.}~\bibnamefont {Hopps}},\ and\ \bibinfo
  {author} {\bibfnamefont {D.}~\bibnamefont {Neely}},\ }\bibfield  {title}
  {\bibinfo {title} {Petawatt class lasers worldwide},\ }\href
  {https://doi.org/10.1017/hpl.2014.52} {\bibfield  {journal} {\bibinfo
  {journal} {High Power Laser Sci. Eng.}\ }\textbf {\bibinfo {volume} {3}},\
  \bibinfo {pages} {e3} (\bibinfo {year} {2015})}\BibitemShut {NoStop}%
\bibitem [{\citenamefont {Danson}\ \emph {et~al.}(2019)\citenamefont {Danson},
  \citenamefont {Haefner}, \citenamefont {Bromage}, \citenamefont {Butcher},
  \citenamefont {Chanteloup}, \citenamefont {Chowdhury}, \citenamefont
  {Galvanauskas}, \citenamefont {Gizzi}, \citenamefont {Hein}, \citenamefont
  {Hillier}, \citenamefont {Hopps}, \citenamefont {Kato}, \citenamefont
  {Khazanov}, \citenamefont {Kodama}, \citenamefont {Korn}, \citenamefont {Li},
  \citenamefont {Li}, \citenamefont {Limpert}, \citenamefont {Ma},
  \citenamefont {Nam}, \citenamefont {Neely}, \citenamefont {Papadopoulos},
  \citenamefont {Penman}, \citenamefont {Qian}, \citenamefont {Rocca},
  \citenamefont {Shaykin}, \citenamefont {Siders}, \citenamefont {Spindloe},
  \citenamefont {Szatmári}, \citenamefont {Trines}, \citenamefont {Zhu},
  \citenamefont {Zhu},\ and\ \citenamefont {Zuegel}}]{Danson2019}%
  \BibitemOpen
  \bibfield  {author} {\bibinfo {author} {\bibfnamefont {C.~N.}\ \bibnamefont
  {Danson}}, \bibinfo {author} {\bibfnamefont {C.}~\bibnamefont {Haefner}},
  \bibinfo {author} {\bibfnamefont {J.}~\bibnamefont {Bromage}}, \bibinfo
  {author} {\bibfnamefont {T.}~\bibnamefont {Butcher}}, \bibinfo {author}
  {\bibfnamefont {J.-C.~F.}\ \bibnamefont {Chanteloup}}, \bibinfo {author}
  {\bibfnamefont {E.~A.}\ \bibnamefont {Chowdhury}}, \bibinfo {author}
  {\bibfnamefont {A.}~\bibnamefont {Galvanauskas}}, \bibinfo {author}
  {\bibfnamefont {L.~A.}\ \bibnamefont {Gizzi}}, \bibinfo {author}
  {\bibfnamefont {J.}~\bibnamefont {Hein}}, \bibinfo {author} {\bibfnamefont
  {D.~I.}\ \bibnamefont {Hillier}}, \bibinfo {author} {\bibfnamefont {N.~W.}\
  \bibnamefont {Hopps}}, \bibinfo {author} {\bibfnamefont {Y.}~\bibnamefont
  {Kato}}, \bibinfo {author} {\bibfnamefont {E.~A.}\ \bibnamefont {Khazanov}},
  \bibinfo {author} {\bibfnamefont {R.}~\bibnamefont {Kodama}}, \bibinfo
  {author} {\bibfnamefont {G.}~\bibnamefont {Korn}}, \bibinfo {author}
  {\bibfnamefont {R.}~\bibnamefont {Li}}, \bibinfo {author} {\bibfnamefont
  {Y.}~\bibnamefont {Li}}, \bibinfo {author} {\bibfnamefont {J.}~\bibnamefont
  {Limpert}}, \bibinfo {author} {\bibfnamefont {J.}~\bibnamefont {Ma}},
  \bibinfo {author} {\bibfnamefont {C.~H.}\ \bibnamefont {Nam}}, \bibinfo
  {author} {\bibfnamefont {D.}~\bibnamefont {Neely}}, \bibinfo {author}
  {\bibfnamefont {D.}~\bibnamefont {Papadopoulos}}, \bibinfo {author}
  {\bibfnamefont {R.~R.}\ \bibnamefont {Penman}}, \bibinfo {author}
  {\bibfnamefont {L.}~\bibnamefont {Qian}}, \bibinfo {author} {\bibfnamefont
  {J.~J.}\ \bibnamefont {Rocca}}, \bibinfo {author} {\bibfnamefont {A.~A.}\
  \bibnamefont {Shaykin}}, \bibinfo {author} {\bibfnamefont {C.~W.}\
  \bibnamefont {Siders}}, \bibinfo {author} {\bibfnamefont {C.}~\bibnamefont
  {Spindloe}}, \bibinfo {author} {\bibfnamefont {S.}~\bibnamefont {Szatmári}},
  \bibinfo {author} {\bibfnamefont {R.~M. G.~M.}\ \bibnamefont {Trines}},
  \bibinfo {author} {\bibfnamefont {J.}~\bibnamefont {Zhu}}, \bibinfo {author}
  {\bibfnamefont {P.}~\bibnamefont {Zhu}},\ and\ \bibinfo {author}
  {\bibfnamefont {J.~D.}\ \bibnamefont {Zuegel}},\ }\bibfield  {title}
  {\bibinfo {title} {Petawatt and exawatt class lasers worldwide},\ }\bibfield
  {journal} {\bibinfo  {journal} {High Power Laser Science and Engineering}\
  }\textbf {\bibinfo {volume} {7}},\ \href
  {https://doi.org/10.1017/hpl.2019.36} {10.1017/hpl.2019.36} (\bibinfo {year}
  {2019})\BibitemShut {NoStop}%
\bibitem [{\citenamefont {{Yoon}}\ \emph {et~al.}(2021)\citenamefont {{Yoon}},
  \citenamefont {{Kim}}, \citenamefont {{Choi}}, \citenamefont {{Sung}},
  \citenamefont {{Lee}}, \citenamefont {{Lee}},\ and\ \citenamefont
  {{Nam}}}]{Yoon_Optica_2021}%
  \BibitemOpen
  \bibfield  {author} {\bibinfo {author} {\bibfnamefont {J.~W.}\ \bibnamefont
  {{Yoon}}}, \bibinfo {author} {\bibfnamefont {Y.~G.}\ \bibnamefont {{Kim}}},
  \bibinfo {author} {\bibfnamefont {I.~W.}\ \bibnamefont {{Choi}}}, \bibinfo
  {author} {\bibfnamefont {J.~H.}\ \bibnamefont {{Sung}}}, \bibinfo {author}
  {\bibfnamefont {H.~W.}\ \bibnamefont {{Lee}}}, \bibinfo {author}
  {\bibfnamefont {S.~K.}\ \bibnamefont {{Lee}}},\ and\ \bibinfo {author}
  {\bibfnamefont {C.~H.}\ \bibnamefont {{Nam}}},\ }\bibfield  {title} {\bibinfo
  {title} {{Realization of laser intensity over 1023 W/cm2}},\ }\href
  {https://doi.org/10.1364/OPTICA.420520} {\bibfield  {journal} {\bibinfo
  {journal} {Optica}\ }\textbf {\bibinfo {volume} {8}},\ \bibinfo {pages} {630}
  (\bibinfo {year} {2021})}\BibitemShut {NoStop}%
\bibitem [{\citenamefont {Radier}\ \emph {et~al.}(2022)\citenamefont {Radier},
  \citenamefont {Chalus}, \citenamefont {Charbonneau}, \citenamefont
  {Thambirajah}, \citenamefont {Deschamps}, \citenamefont {David},
  \citenamefont {Barbe}, \citenamefont {Etter}, \citenamefont {Matras},
  \citenamefont {Ricaud},\ and\ \citenamefont {et~al.}}]{Radier_HPLSE_2022}%
  \BibitemOpen
  \bibfield  {author} {\bibinfo {author} {\bibfnamefont {C.}~\bibnamefont
  {Radier}}, \bibinfo {author} {\bibfnamefont {O.}~\bibnamefont {Chalus}},
  \bibinfo {author} {\bibfnamefont {M.}~\bibnamefont {Charbonneau}}, \bibinfo
  {author} {\bibfnamefont {S.}~\bibnamefont {Thambirajah}}, \bibinfo {author}
  {\bibfnamefont {G.}~\bibnamefont {Deschamps}}, \bibinfo {author}
  {\bibfnamefont {S.}~\bibnamefont {David}}, \bibinfo {author} {\bibfnamefont
  {J.}~\bibnamefont {Barbe}}, \bibinfo {author} {\bibfnamefont
  {E.}~\bibnamefont {Etter}}, \bibinfo {author} {\bibfnamefont
  {G.}~\bibnamefont {Matras}}, \bibinfo {author} {\bibfnamefont
  {S.}~\bibnamefont {Ricaud}},\ and\ \bibinfo {author} {\bibnamefont
  {et~al.}},\ }\bibfield  {title} {\bibinfo {title} {{10 PW peak power
  femtosecond laser pulses at ELI-NP}},\ }\href
  {https://doi.org/10.1017/hpl.2022.11} {\bibfield  {journal} {\bibinfo
  {journal} {{High Power Laser Sci. Eng.}}\ }\textbf {\bibinfo {volume} {10}},\
  \bibinfo {pages} {e21} (\bibinfo {year} {2022})}\BibitemShut {NoStop}%
\bibitem [{\citenamefont {{Nikishov}}\ and\ \citenamefont
  {{Ritus}}(1964)}]{Nikishov_JETP_1964}%
  \BibitemOpen
  \bibfield  {author} {\bibinfo {author} {\bibfnamefont {A.~I.}\ \bibnamefont
  {{Nikishov}}}\ and\ \bibinfo {author} {\bibfnamefont {V.~I.}\ \bibnamefont
  {{Ritus}}},\ }\bibfield  {title} {\bibinfo {title} {{Pair Production by a
  Photon and Photon Emission by an Electron in the Field of an Intense
  Electromagnetic Wave and in a Constant Field}},\ }\href
  {http://jetp.ac.ru/cgi-bin/ dn/e_019_02_0529.pdf} {\bibfield  {journal}
  {\bibinfo  {journal} {{Sov. Phys. JETP}}\ }\textbf {\bibinfo {volume} {19}},\
  \bibinfo {pages} {529} (\bibinfo {year} {1964})}\BibitemShut {NoStop}%
\bibitem [{\citenamefont {{Erber}}(1966)}]{Erber_RMP_1966}%
  \BibitemOpen
  \bibfield  {author} {\bibinfo {author} {\bibfnamefont {T.}~\bibnamefont
  {{Erber}}},\ }\bibfield  {title} {\bibinfo {title} {{High-Energy
  Electromagnetic Conversion Processes in Intense Magnetic Fields}},\ }\href
  {https://doi.org/10.1103/RevModPhys.38.626} {\bibfield  {journal} {\bibinfo
  {journal} {Rev. Mod. Phys.}\ }\textbf {\bibinfo {volume} {38}},\ \bibinfo
  {pages} {626} (\bibinfo {year} {1966})}\BibitemShut {NoStop}%
\bibitem [{\citenamefont {{Di Piazza}}\ \emph {et~al.}(2012)\citenamefont {{Di
  Piazza}}, \citenamefont {{M{\"u}ller}}, \citenamefont {{Hatsagortsyan}},\
  and\ \citenamefont {{Keitel}}}]{DiPiazza_RMP_2012}%
  \BibitemOpen
  \bibfield  {author} {\bibinfo {author} {\bibfnamefont {A.}~\bibnamefont {{Di
  Piazza}}}, \bibinfo {author} {\bibfnamefont {C.}~\bibnamefont
  {{M{\"u}ller}}}, \bibinfo {author} {\bibfnamefont {K.~Z.}\ \bibnamefont
  {{Hatsagortsyan}}},\ and\ \bibinfo {author} {\bibfnamefont {C.~H.}\
  \bibnamefont {{Keitel}}},\ }\bibfield  {title} {\bibinfo {title} {{Extremely
  high-intensity laser interactions with fundamental quantum systems}},\ }\href
  {https://doi.org/10.1103/RevModPhys.84.1177} {\bibfield  {journal} {\bibinfo
  {journal} {Rev. Mod. Phys.}\ }\textbf {\bibinfo {volume} {84}},\ \bibinfo
  {pages} {1177} (\bibinfo {year} {2012})}\BibitemShut {NoStop}%
\bibitem [{\citenamefont {{Bell}}\ and\ \citenamefont
  {{Kirk}}(2008)}]{Bell_PRL_2008}%
  \BibitemOpen
  \bibfield  {author} {\bibinfo {author} {\bibfnamefont {A.~R.}\ \bibnamefont
  {{Bell}}}\ and\ \bibinfo {author} {\bibfnamefont {J.~G.}\ \bibnamefont
  {{Kirk}}},\ }\bibfield  {title} {\bibinfo {title} {{Possibility of Prolific
  Pair Production with High-Power Lasers}},\ }\href
  {https://doi.org/10.1103/PhysRevLett.101.200403} {\bibfield  {journal}
  {\bibinfo  {journal} {Phys. Rev. Lett.}\ }\textbf {\bibinfo {volume} {101}},\
  \bibinfo {eid} {200403} (\bibinfo {year} {2008})}\BibitemShut {NoStop}%
\bibitem [{\citenamefont {Ridgers}\ \emph {et~al.}(2012)\citenamefont
  {Ridgers}, \citenamefont {Brady}, \citenamefont {Duclous}, \citenamefont
  {Kirk}, \citenamefont {Bennett}, \citenamefont {Arber}, \citenamefont
  {Robinson},\ and\ \citenamefont {Bell}}]{Ridgers_PRL_2012}%
  \BibitemOpen
  \bibfield  {author} {\bibinfo {author} {\bibfnamefont {C.~P.}\ \bibnamefont
  {Ridgers}}, \bibinfo {author} {\bibfnamefont {C.~S.}\ \bibnamefont {Brady}},
  \bibinfo {author} {\bibfnamefont {R.}~\bibnamefont {Duclous}}, \bibinfo
  {author} {\bibfnamefont {J.~G.}\ \bibnamefont {Kirk}}, \bibinfo {author}
  {\bibfnamefont {K.}~\bibnamefont {Bennett}}, \bibinfo {author} {\bibfnamefont
  {T.~D.}\ \bibnamefont {Arber}}, \bibinfo {author} {\bibfnamefont {A.~P.~L.}\
  \bibnamefont {Robinson}},\ and\ \bibinfo {author} {\bibfnamefont {A.~R.}\
  \bibnamefont {Bell}},\ }\bibfield  {title} {\bibinfo {title} {Dense
  electron-positron plasmas and ultraintense $\ensuremath{\gamma}$ rays from
  laser-irradiated solids},\ }\href
  {https://doi.org/10.1103/PhysRevLett.108.165006} {\bibfield  {journal}
  {\bibinfo  {journal} {Phys. Rev. Lett.}\ }\textbf {\bibinfo {volume} {108}},\
  \bibinfo {pages} {165006} (\bibinfo {year} {2012})}\BibitemShut {NoStop}%
\bibitem [{\citenamefont {Vranic}\ \emph {et~al.}(2014)\citenamefont {Vranic},
  \citenamefont {Martins}, \citenamefont {Vieira}, \citenamefont {Fonseca},\
  and\ \citenamefont {Silva}}]{Vranic_PRL_2014}%
  \BibitemOpen
  \bibfield  {author} {\bibinfo {author} {\bibfnamefont {M.}~\bibnamefont
  {Vranic}}, \bibinfo {author} {\bibfnamefont {J.~L.}\ \bibnamefont {Martins}},
  \bibinfo {author} {\bibfnamefont {J.}~\bibnamefont {Vieira}}, \bibinfo
  {author} {\bibfnamefont {R.~A.}\ \bibnamefont {Fonseca}},\ and\ \bibinfo
  {author} {\bibfnamefont {L.~O.}\ \bibnamefont {Silva}},\ }\bibfield  {title}
  {\bibinfo {title} {All-optical radiation reaction at $1{0}^{21}\text{ }\text{
  }\mathrm{W}/{\mathrm{cm}}^{2}$},\ }\href
  {https://doi.org/10.1103/PhysRevLett.113.134801} {\bibfield  {journal}
  {\bibinfo  {journal} {Phys. Rev. Lett.}\ }\textbf {\bibinfo {volume} {113}},\
  \bibinfo {pages} {134801} (\bibinfo {year} {2014})}\BibitemShut {NoStop}%
\bibitem [{\citenamefont {Lobet}\ \emph {et~al.}(2017)\citenamefont {Lobet},
  \citenamefont {Davoine}, \citenamefont {d'Humi\`eres},\ and\ \citenamefont
  {Gremillet}}]{Lobet_PRAB_2017}%
  \BibitemOpen
  \bibfield  {author} {\bibinfo {author} {\bibfnamefont {M.}~\bibnamefont
  {Lobet}}, \bibinfo {author} {\bibfnamefont {X.}~\bibnamefont {Davoine}},
  \bibinfo {author} {\bibfnamefont {E.}~\bibnamefont {d'Humi\`eres}},\ and\
  \bibinfo {author} {\bibfnamefont {L.}~\bibnamefont {Gremillet}},\ }\bibfield
  {title} {\bibinfo {title} {Generation of high-energy electron-positron pairs
  in the collision of a laser-accelerated electron beam with a multipetawatt
  laser},\ }\href {https://doi.org/10.1103/PhysRevAccelBeams.20.043401}
  {\bibfield  {journal} {\bibinfo  {journal} {Phys. Rev. Accel. Beams}\
  }\textbf {\bibinfo {volume} {20}},\ \bibinfo {pages} {043401} (\bibinfo
  {year} {2017})}\BibitemShut {NoStop}%
\bibitem [{\citenamefont {{Magnusson}}\ \emph {et~al.}(2019)\citenamefont
  {{Magnusson}}, \citenamefont {{Gonoskov}}, \citenamefont {{Marklund}},
  \citenamefont {{Esirkepov}}, \citenamefont {{Koga}}, \citenamefont {{Kondo}},
  \citenamefont {{Kando}}, \citenamefont {{Bulanov}}, \citenamefont {{Korn}},\
  and\ \citenamefont {{Bulanov}}}]{Magnusson_PRL_2019}%
  \BibitemOpen
  \bibfield  {author} {\bibinfo {author} {\bibfnamefont {J.}~\bibnamefont
  {{Magnusson}}}, \bibinfo {author} {\bibfnamefont {A.}~\bibnamefont
  {{Gonoskov}}}, \bibinfo {author} {\bibfnamefont {M.}~\bibnamefont
  {{Marklund}}}, \bibinfo {author} {\bibfnamefont {T.~Z.}\ \bibnamefont
  {{Esirkepov}}}, \bibinfo {author} {\bibfnamefont {J.~K.}\ \bibnamefont
  {{Koga}}}, \bibinfo {author} {\bibfnamefont {K.}~\bibnamefont {{Kondo}}},
  \bibinfo {author} {\bibfnamefont {M.}~\bibnamefont {{Kando}}}, \bibinfo
  {author} {\bibfnamefont {S.~V.}\ \bibnamefont {{Bulanov}}}, \bibinfo {author}
  {\bibfnamefont {G.}~\bibnamefont {{Korn}}},\ and\ \bibinfo {author}
  {\bibfnamefont {S.~S.}\ \bibnamefont {{Bulanov}}},\ }\bibfield  {title}
  {\bibinfo {title} {{Laser-Particle Collider for Multi-GeV Photon
  Production}},\ }\href {https://doi.org/10.1103/PhysRevLett.122.254801}
  {\bibfield  {journal} {\bibinfo  {journal} {Phys. Rev. Lett.}\ }\textbf
  {\bibinfo {volume} {122}},\ \bibinfo {eid} {254801} (\bibinfo {year}
  {2019})}\BibitemShut {NoStop}%
\bibitem [{\citenamefont {{Vincenti}}(2019)}]{Vincenti_PRL_2019}%
  \BibitemOpen
  \bibfield  {author} {\bibinfo {author} {\bibfnamefont {H.}~\bibnamefont
  {{Vincenti}}},\ }\bibfield  {title} {\bibinfo {title} {{Achieving Extreme
  Light Intensities using Optically Curved Relativistic Plasma Mirrors}},\
  }\href {https://doi.org/10.1103/PhysRevLett.123.105001} {\bibfield  {journal}
  {\bibinfo  {journal} {Phys. Rev. Lett.}\ }\textbf {\bibinfo {volume} {123}},\
  \bibinfo {eid} {105001} (\bibinfo {year} {2019})}\BibitemShut {NoStop}%
\bibitem [{\citenamefont {{L{\'e}cz}}\ and\ \citenamefont
  {{Andreev}}(2019)}]{Lecz_PPCF_2019}%
  \BibitemOpen
  \bibfield  {author} {\bibinfo {author} {\bibfnamefont {Z.}~\bibnamefont
  {{L{\'e}cz}}}\ and\ \bibinfo {author} {\bibfnamefont {A.}~\bibnamefont
  {{Andreev}}},\ }\bibfield  {title} {\bibinfo {title} {{Minimum requirements
  for electron-positron pair creation in the interaction of ultra-short laser
  pulses with thin foils}},\ }\href {https://doi.org/10.1088/1361-6587/aafe59}
  {\bibfield  {journal} {\bibinfo  {journal} {Plasma Phys. Control. Fusion}\
  }\textbf {\bibinfo {volume} {61}},\ \bibinfo {eid} {045005} (\bibinfo {year}
  {2019})}\BibitemShut {NoStop}%
\bibitem [{\citenamefont {Blackburn}(2020)}]{Blackburn_RMPP_2020}%
  \BibitemOpen
  \bibfield  {author} {\bibinfo {author} {\bibfnamefont {T.~G.}\ \bibnamefont
  {Blackburn}},\ }\bibfield  {title} {\bibinfo {title} {{Radiation reaction in
  electron–beam interactions with high-intensity lasers}},\ }\href
  {https://doi.org/10.1007/s41614-020-0042-0} {\bibfield  {journal} {\bibinfo
  {journal} {Rev. Mod. Plasma Phys.}\ }\textbf {\bibinfo {volume} {4}},\
  \bibinfo {eid} {5} (\bibinfo {year} {2020})}\BibitemShut {NoStop}%
\bibitem [{\citenamefont {Mercuri-Baron}\ \emph {et~al.}(2021)\citenamefont
  {Mercuri-Baron}, \citenamefont {Grech}, \citenamefont {Niel}, \citenamefont
  {Grassi}, \citenamefont {Lobet}, \citenamefont {Piazza},\ and\ \citenamefont
  {Riconda}}]{Mercuri_Baron_2021}%
  \BibitemOpen
  \bibfield  {author} {\bibinfo {author} {\bibfnamefont {A.}~\bibnamefont
  {Mercuri-Baron}}, \bibinfo {author} {\bibfnamefont {M.}~\bibnamefont
  {Grech}}, \bibinfo {author} {\bibfnamefont {F.}~\bibnamefont {Niel}},
  \bibinfo {author} {\bibfnamefont {A.}~\bibnamefont {Grassi}}, \bibinfo
  {author} {\bibfnamefont {M.}~\bibnamefont {Lobet}}, \bibinfo {author}
  {\bibfnamefont {A.~D.}\ \bibnamefont {Piazza}},\ and\ \bibinfo {author}
  {\bibfnamefont {C.}~\bibnamefont {Riconda}},\ }\bibfield  {title} {\bibinfo
  {title} {{Impact of the laser spatio-temporal shape on
  Breit{\textendash}Wheeler pair production}},\ }\href
  {https://doi.org/10.1088/1367-2630/ac1975} {\bibfield  {journal} {\bibinfo
  {journal} {New J. Phys.}\ }\textbf {\bibinfo {volume} {23}},\ \bibinfo
  {pages} {085006} (\bibinfo {year} {2021})}\BibitemShut {NoStop}%
\bibitem [{\citenamefont {{Qu}}\ \emph {et~al.}(2021)\citenamefont {{Qu}},
  \citenamefont {{Meuren}},\ and\ \citenamefont {{Fisch}}}]{Qu_PRL_2021}%
  \BibitemOpen
  \bibfield  {author} {\bibinfo {author} {\bibfnamefont {K.}~\bibnamefont
  {{Qu}}}, \bibinfo {author} {\bibfnamefont {S.}~\bibnamefont {{Meuren}}},\
  and\ \bibinfo {author} {\bibfnamefont {N.~J.}\ \bibnamefont {{Fisch}}},\
  }\bibfield  {title} {\bibinfo {title} {{Signature of Collective Plasma
  Effects in Beam-Driven QED Cascades}},\ }\href
  {https://doi.org/10.1103/PhysRevLett.127.095001} {\bibfield  {journal}
  {\bibinfo  {journal} {Phys. Rev. Lett.}\ }\textbf {\bibinfo {volume} {127}},\
  \bibinfo {eid} {095001} (\bibinfo {year} {2021})}\BibitemShut {NoStop}%
\bibitem [{\citenamefont {Di~Piazza}\ \emph {et~al.}(2020)\citenamefont
  {Di~Piazza}, \citenamefont {Wistisen}, \citenamefont {Tamburini},\ and\
  \citenamefont {Uggerh\o{}j}}]{PhysRevLett.124.044801}%
  \BibitemOpen
  \bibfield  {author} {\bibinfo {author} {\bibfnamefont {A.}~\bibnamefont
  {Di~Piazza}}, \bibinfo {author} {\bibfnamefont {T.~N.}\ \bibnamefont
  {Wistisen}}, \bibinfo {author} {\bibfnamefont {M.}~\bibnamefont
  {Tamburini}},\ and\ \bibinfo {author} {\bibfnamefont {U.~I.}\ \bibnamefont
  {Uggerh\o{}j}},\ }\bibfield  {title} {\bibinfo {title} {Testing strong field
  qed close to the fully nonperturbative regime using aligned crystals},\
  }\href {https://doi.org/10.1103/PhysRevLett.124.044801} {\bibfield  {journal}
  {\bibinfo  {journal} {Phys. Rev. Lett.}\ }\textbf {\bibinfo {volume} {124}},\
  \bibinfo {pages} {044801} (\bibinfo {year} {2020})}\BibitemShut {NoStop}%
\bibitem [{\citenamefont {Baumann}\ \emph {et~al.}(2019)\citenamefont
  {Baumann}, \citenamefont {Nerush}, \citenamefont {Pukhov},\ and\
  \citenamefont {Kostyukov}}]{BaumannQED}%
  \BibitemOpen
  \bibfield  {author} {\bibinfo {author} {\bibfnamefont {C.}~\bibnamefont
  {Baumann}}, \bibinfo {author} {\bibfnamefont {E.~N.}\ \bibnamefont {Nerush}},
  \bibinfo {author} {\bibfnamefont {A.}~\bibnamefont {Pukhov}},\ and\ \bibinfo
  {author} {\bibfnamefont {I.~Y.}\ \bibnamefont {Kostyukov}},\ }\bibfield
  {title} {\bibinfo {title} {Probing non-perturbative qed with electron-laser
  collisions},\ }\href {https://doi.org/10.1038/s41598-019-45582-5} {\bibfield
  {journal} {\bibinfo  {journal} {Scientific Reports}\ }\textbf {\bibinfo
  {volume} {9}},\ \bibinfo {pages} {9407} (\bibinfo {year} {2019})}\BibitemShut
  {NoStop}%
\bibitem [{\citenamefont {Schwinger}(1951)}]{Schwinger_PR_1951}%
  \BibitemOpen
  \bibfield  {author} {\bibinfo {author} {\bibfnamefont {J.}~\bibnamefont
  {Schwinger}},\ }\bibfield  {title} {\bibinfo {title} {On gauge invariance and
  vacuum polarization},\ }\href {https://doi.org/10.1103/PhysRev.82.664}
  {\bibfield  {journal} {\bibinfo  {journal} {Phys. Rev.}\ }\textbf {\bibinfo
  {volume} {82}},\ \bibinfo {pages} {664} (\bibinfo {year} {1951})}\BibitemShut
  {NoStop}%
\bibitem [{\citenamefont {Wolkow}(1935)}]{Volkov1935}%
  \BibitemOpen
  \bibfield  {author} {\bibinfo {author} {\bibfnamefont {D.~M.}\ \bibnamefont
  {Wolkow}},\ }\bibfield  {title} {\bibinfo {title} {ber eine klasse von
  lsungen der diracschen gleichung},\ }\href
  {https://doi.org/10.1007/bf01331022} {\bibfield  {journal} {\bibinfo
  {journal} {Zeitschrift fr Physik}\ }\textbf {\bibinfo {volume} {94}},\
  \bibinfo {pages} {250–260} (\bibinfo {year} {1935})}\BibitemShut {NoStop}%
\bibitem [{\citenamefont {Berestetskii}\ \emph {et~al.}(1982)\citenamefont
  {Berestetskii}, \citenamefont {Lifshitz},\ and\ \citenamefont
  {Pitaevskii}}]{Berestetskii1982}%
  \BibitemOpen
  \bibfield  {author} {\bibinfo {author} {\bibfnamefont {V.~B.}\ \bibnamefont
  {Berestetskii}}, \bibinfo {author} {\bibfnamefont {E.~M.}\ \bibnamefont
  {Lifshitz}},\ and\ \bibinfo {author} {\bibfnamefont {L.~P.}\ \bibnamefont
  {Pitaevskii}},\ }\href {https://doi.org/10.1016/C2009-0-24486-2} {\emph
  {\bibinfo {title} {Quantum Electrodynamics (Second Edition)}}}\ (\bibinfo
  {publisher} {Butterworth-Heinemann},\ \bibinfo {address} {Berlin},\ \bibinfo
  {year} {1982})\BibitemShut {NoStop}%
\bibitem [{\citenamefont {{Kirk}}\ \emph {et~al.}(2009)\citenamefont {{Kirk}},
  \citenamefont {{Bell}},\ and\ \citenamefont {{Arka}}}]{Kirk_PPCF_2009}%
  \BibitemOpen
  \bibfield  {author} {\bibinfo {author} {\bibfnamefont {J.~G.}\ \bibnamefont
  {{Kirk}}}, \bibinfo {author} {\bibfnamefont {A.~R.}\ \bibnamefont {{Bell}}},\
  and\ \bibinfo {author} {\bibfnamefont {I.}~\bibnamefont {{Arka}}},\
  }\bibfield  {title} {\bibinfo {title} {{Pair production in
  counter-propagating laser beams}},\ }\href
  {https://doi.org/10.1088/0741-3335/51/8/085008} {\bibfield  {journal}
  {\bibinfo  {journal} {Plasma Phys. Control. Fusion}\ }\textbf {\bibinfo
  {volume} {51}},\ \bibinfo {eid} {085008} (\bibinfo {year}
  {2009})}\BibitemShut {NoStop}%
\bibitem [{\citenamefont {{Ritus}}(1972)}]{Ritus_AP_1972}%
  \BibitemOpen
  \bibfield  {author} {\bibinfo {author} {\bibfnamefont {V.~I.}\ \bibnamefont
  {{Ritus}}},\ }\bibfield  {title} {\bibinfo {title} {{Radiative corrections in
  quantum electrodynamics with intense field and their analytical
  properties}},\ }\href {https://doi.org/10.1016/0003-4916(72)90191-1}
  {\bibfield  {journal} {\bibinfo  {journal} {Ann. Phys.}\ }\textbf {\bibinfo
  {volume} {69}},\ \bibinfo {pages} {555} (\bibinfo {year} {1972})}\BibitemShut
  {NoStop}%
\bibitem [{\citenamefont {Narozhny}(1980)}]{Narozhny_PRD_1980}%
  \BibitemOpen
  \bibfield  {author} {\bibinfo {author} {\bibfnamefont {N.~B.}\ \bibnamefont
  {Narozhny}},\ }\bibfield  {title} {\bibinfo {title} {Expansion parameter of
  perturbation theory in intense-field quantum electrodynamics},\ }\href
  {https://doi.org/10.1103/PhysRevD.21.1176} {\bibfield  {journal} {\bibinfo
  {journal} {Phys. Rev. D}\ }\textbf {\bibinfo {volume} {21}},\ \bibinfo
  {pages} {1176} (\bibinfo {year} {1980})}\BibitemShut {NoStop}%
\bibitem [{\citenamefont {{Fedotov}}(2017)}]{Fedotov_JPhCS_2017}%
  \BibitemOpen
  \bibfield  {author} {\bibinfo {author} {\bibfnamefont {A.}~\bibnamefont
  {{Fedotov}}},\ }\bibfield  {title} {\bibinfo {title} {{Conjecture of
  perturbative QED breakdown at
  {\ensuremath{\alpha}}{\ensuremath{\chi}}$^{2/3}$ {\ensuremath{\gtrsim}} 1}},\
  }\href {https://doi.org/10.1088/1742-6596/826/1/012027} {\bibfield  {journal}
  {\bibinfo  {journal} {J. Phys. Conf. Ser.}\ }\textbf {\bibinfo {volume}
  {826}},\ \bibinfo {eid} {012027} (\bibinfo {year} {2017})}\BibitemShut
  {NoStop}%
\bibitem [{\citenamefont {Fedotov}\ \emph {et~al.}(2023)\citenamefont
  {Fedotov}, \citenamefont {Ilderton}, \citenamefont {Karbstein}, \citenamefont
  {King}, \citenamefont {Seipt}, \citenamefont {Taya},\ and\ \citenamefont
  {Torgrimsson}}]{Fedotov2023}%
  \BibitemOpen
  \bibfield  {author} {\bibinfo {author} {\bibfnamefont {A.}~\bibnamefont
  {Fedotov}}, \bibinfo {author} {\bibfnamefont {A.}~\bibnamefont {Ilderton}},
  \bibinfo {author} {\bibfnamefont {F.}~\bibnamefont {Karbstein}}, \bibinfo
  {author} {\bibfnamefont {B.}~\bibnamefont {King}}, \bibinfo {author}
  {\bibfnamefont {D.}~\bibnamefont {Seipt}}, \bibinfo {author} {\bibfnamefont
  {H.}~\bibnamefont {Taya}},\ and\ \bibinfo {author} {\bibfnamefont
  {G.}~\bibnamefont {Torgrimsson}},\ }\bibfield  {title} {\bibinfo {title}
  {Advances in qed with intense background fields},\ }\href
  {https://doi.org/10.1016/j.physrep.2023.01.003} {\bibfield  {journal}
  {\bibinfo  {journal} {Phys. Rep.}\ }\textbf {\bibinfo {volume} {1010}},\
  \bibinfo {pages} {1–138} (\bibinfo {year} {2023})}\BibitemShut {NoStop}%
\bibitem [{\citenamefont {Burke}\ \emph {et~al.}(1997)\citenamefont {Burke},
  \citenamefont {Field}, \citenamefont {Horton-Smith}, \citenamefont {Spencer},
  \citenamefont {Walz}, \citenamefont {Berridge}, \citenamefont {Bugg},
  \citenamefont {Shmakov}, \citenamefont {Weidemann}, \citenamefont {Bula},
  \citenamefont {McDonald}, \citenamefont {Prebys}, \citenamefont {Bamber},
  \citenamefont {Boege}, \citenamefont {Koffas}, \citenamefont {Kotseroglou},
  \citenamefont {Melissinos}, \citenamefont {Meyerhofer}, \citenamefont
  {Reis},\ and\ \citenamefont {Ragg}}]{Burke_PRL_1997}%
  \BibitemOpen
  \bibfield  {author} {\bibinfo {author} {\bibfnamefont {D.~L.}\ \bibnamefont
  {Burke}}, \bibinfo {author} {\bibfnamefont {R.~C.}\ \bibnamefont {Field}},
  \bibinfo {author} {\bibfnamefont {G.}~\bibnamefont {Horton-Smith}}, \bibinfo
  {author} {\bibfnamefont {J.~E.}\ \bibnamefont {Spencer}}, \bibinfo {author}
  {\bibfnamefont {D.}~\bibnamefont {Walz}}, \bibinfo {author} {\bibfnamefont
  {S.~C.}\ \bibnamefont {Berridge}}, \bibinfo {author} {\bibfnamefont {W.~M.}\
  \bibnamefont {Bugg}}, \bibinfo {author} {\bibfnamefont {K.}~\bibnamefont
  {Shmakov}}, \bibinfo {author} {\bibfnamefont {A.~W.}\ \bibnamefont
  {Weidemann}}, \bibinfo {author} {\bibfnamefont {C.}~\bibnamefont {Bula}},
  \bibinfo {author} {\bibfnamefont {K.~T.}\ \bibnamefont {McDonald}}, \bibinfo
  {author} {\bibfnamefont {E.~J.}\ \bibnamefont {Prebys}}, \bibinfo {author}
  {\bibfnamefont {C.}~\bibnamefont {Bamber}}, \bibinfo {author} {\bibfnamefont
  {S.~J.}\ \bibnamefont {Boege}}, \bibinfo {author} {\bibfnamefont
  {T.}~\bibnamefont {Koffas}}, \bibinfo {author} {\bibfnamefont
  {T.}~\bibnamefont {Kotseroglou}}, \bibinfo {author} {\bibfnamefont {A.~C.}\
  \bibnamefont {Melissinos}}, \bibinfo {author} {\bibfnamefont {D.~D.}\
  \bibnamefont {Meyerhofer}}, \bibinfo {author} {\bibfnamefont {D.~A.}\
  \bibnamefont {Reis}},\ and\ \bibinfo {author} {\bibfnamefont
  {W.}~\bibnamefont {Ragg}},\ }\bibfield  {title} {\bibinfo {title} {Positron
  production in multiphoton light-by-light scattering},\ }\href
  {https://doi.org/10.1103/PhysRevLett.79.1626} {\bibfield  {journal} {\bibinfo
   {journal} {Phys. Rev. Lett.}\ }\textbf {\bibinfo {volume} {79}},\ \bibinfo
  {pages} {1626} (\bibinfo {year} {1997})}\BibitemShut {NoStop}%
\bibitem [{\citenamefont {Bamber}\ \emph {et~al.}(1999)\citenamefont {Bamber},
  \citenamefont {Boege}, \citenamefont {Koffas}, \citenamefont {Kotseroglou},
  \citenamefont {Melissinos}, \citenamefont {Meyerhofer}, \citenamefont {Reis},
  \citenamefont {Ragg}, \citenamefont {Bula}, \citenamefont {McDonald},
  \citenamefont {Prebys}, \citenamefont {Burke}, \citenamefont {Field},
  \citenamefont {Horton-Smith}, \citenamefont {Spencer}, \citenamefont {Walz},
  \citenamefont {Berridge}, \citenamefont {Bugg}, \citenamefont {Shmakov},\
  and\ \citenamefont {Weidemann}}]{Bamber_PRD_1999}%
  \BibitemOpen
  \bibfield  {author} {\bibinfo {author} {\bibfnamefont {C.}~\bibnamefont
  {Bamber}}, \bibinfo {author} {\bibfnamefont {S.~J.}\ \bibnamefont {Boege}},
  \bibinfo {author} {\bibfnamefont {T.}~\bibnamefont {Koffas}}, \bibinfo
  {author} {\bibfnamefont {T.}~\bibnamefont {Kotseroglou}}, \bibinfo {author}
  {\bibfnamefont {A.~C.}\ \bibnamefont {Melissinos}}, \bibinfo {author}
  {\bibfnamefont {D.~D.}\ \bibnamefont {Meyerhofer}}, \bibinfo {author}
  {\bibfnamefont {D.~A.}\ \bibnamefont {Reis}}, \bibinfo {author}
  {\bibfnamefont {W.}~\bibnamefont {Ragg}}, \bibinfo {author} {\bibfnamefont
  {C.}~\bibnamefont {Bula}}, \bibinfo {author} {\bibfnamefont {K.~T.}\
  \bibnamefont {McDonald}}, \bibinfo {author} {\bibfnamefont {E.~J.}\
  \bibnamefont {Prebys}}, \bibinfo {author} {\bibfnamefont {D.~L.}\
  \bibnamefont {Burke}}, \bibinfo {author} {\bibfnamefont {R.~C.}\ \bibnamefont
  {Field}}, \bibinfo {author} {\bibfnamefont {G.}~\bibnamefont {Horton-Smith}},
  \bibinfo {author} {\bibfnamefont {J.~E.}\ \bibnamefont {Spencer}}, \bibinfo
  {author} {\bibfnamefont {D.}~\bibnamefont {Walz}}, \bibinfo {author}
  {\bibfnamefont {S.~C.}\ \bibnamefont {Berridge}}, \bibinfo {author}
  {\bibfnamefont {W.~M.}\ \bibnamefont {Bugg}}, \bibinfo {author}
  {\bibfnamefont {K.}~\bibnamefont {Shmakov}},\ and\ \bibinfo {author}
  {\bibfnamefont {A.~W.}\ \bibnamefont {Weidemann}},\ }\bibfield  {title}
  {\bibinfo {title} {Studies of nonlinear qed in collisions of 46.6 gev
  electrons with intense laser pulses},\ }\href
  {https://doi.org/10.1103/PhysRevD.60.092004} {\bibfield  {journal} {\bibinfo
  {journal} {Phys. Rev. D}\ }\textbf {\bibinfo {volume} {60}},\ \bibinfo
  {pages} {092004} (\bibinfo {year} {1999})}\BibitemShut {NoStop}%
\bibitem [{\citenamefont {Tajima}\ and\ \citenamefont
  {Dawson}(1979)}]{Tajima_PRL_1979}%
  \BibitemOpen
  \bibfield  {author} {\bibinfo {author} {\bibfnamefont {T.}~\bibnamefont
  {Tajima}}\ and\ \bibinfo {author} {\bibfnamefont {J.~M.}\ \bibnamefont
  {Dawson}},\ }\bibfield  {title} {\bibinfo {title} {Laser electron
  accelerator},\ }\href {https://doi.org/10.1103/PhysRevLett.43.267} {\bibfield
   {journal} {\bibinfo  {journal} {Phys. Rev. Lett.}\ }\textbf {\bibinfo
  {volume} {43}},\ \bibinfo {pages} {267} (\bibinfo {year} {1979})}\BibitemShut
  {NoStop}%
\bibitem [{\citenamefont {Faure}\ \emph {et~al.}(2004)\citenamefont {Faure},
  \citenamefont {Glinec}, \citenamefont {Pukhov}, \citenamefont {Kiselev},
  \citenamefont {Gordienko}, \citenamefont {Lefebvre}, \citenamefont
  {Rousseau}, \citenamefont {Burgy},\ and\ \citenamefont
  {Malka}}]{Faure_Nature_2004}%
  \BibitemOpen
  \bibfield  {author} {\bibinfo {author} {\bibfnamefont {J.}~\bibnamefont
  {Faure}}, \bibinfo {author} {\bibfnamefont {Y.}~\bibnamefont {Glinec}},
  \bibinfo {author} {\bibfnamefont {A.}~\bibnamefont {Pukhov}}, \bibinfo
  {author} {\bibfnamefont {S.}~\bibnamefont {Kiselev}}, \bibinfo {author}
  {\bibfnamefont {S.}~\bibnamefont {Gordienko}}, \bibinfo {author}
  {\bibfnamefont {E.}~\bibnamefont {Lefebvre}}, \bibinfo {author}
  {\bibfnamefont {J.-P.}\ \bibnamefont {Rousseau}}, \bibinfo {author}
  {\bibfnamefont {F.}~\bibnamefont {Burgy}},\ and\ \bibinfo {author}
  {\bibfnamefont {V.}~\bibnamefont {Malka}},\ }\bibfield  {title} {\bibinfo
  {title} {A laser--plasma accelerator producing monoenergetic electron
  beams},\ }\href {https://doi.org/10.1038/nature02963} {\bibfield  {journal}
  {\bibinfo  {journal} {Nature}\ }\textbf {\bibinfo {volume} {431}},\ \bibinfo
  {pages} {541} (\bibinfo {year} {2004})}\BibitemShut {NoStop}%
\bibitem [{\citenamefont {Geddes}\ \emph {et~al.}(2004)\citenamefont {Geddes},
  \citenamefont {Toth}, \citenamefont {van Tilborg}, \citenamefont {Esarey},
  \citenamefont {Schroeder}, \citenamefont {Bruhwiler}, \citenamefont {Nieter},
  \citenamefont {Cary},\ and\ \citenamefont {Leemans}}]{Geddes_Nature_2004}%
  \BibitemOpen
  \bibfield  {author} {\bibinfo {author} {\bibfnamefont {C.~G.~R.}\
  \bibnamefont {Geddes}}, \bibinfo {author} {\bibfnamefont {C.}~\bibnamefont
  {Toth}}, \bibinfo {author} {\bibfnamefont {J.}~\bibnamefont {van Tilborg}},
  \bibinfo {author} {\bibfnamefont {E.}~\bibnamefont {Esarey}}, \bibinfo
  {author} {\bibfnamefont {C.~B.}\ \bibnamefont {Schroeder}}, \bibinfo {author}
  {\bibfnamefont {D.}~\bibnamefont {Bruhwiler}}, \bibinfo {author}
  {\bibfnamefont {C.}~\bibnamefont {Nieter}}, \bibinfo {author} {\bibfnamefont
  {J.}~\bibnamefont {Cary}},\ and\ \bibinfo {author} {\bibfnamefont {W.~P.}\
  \bibnamefont {Leemans}},\ }\bibfield  {title} {\bibinfo {title} {High-quality
  electron beams from a laser wakefield accelerator using plasma-channel
  guiding},\ }\href {https://doi.org/10.1038/nature02900} {\bibfield  {journal}
  {\bibinfo  {journal} {Nature}\ }\textbf {\bibinfo {volume} {431}},\ \bibinfo
  {pages} {538} (\bibinfo {year} {2004})}\BibitemShut {NoStop}%
\bibitem [{\citenamefont {Mangles}\ \emph {et~al.}(2004)\citenamefont
  {Mangles}, \citenamefont {Murphy}, \citenamefont {Najmudin}, \citenamefont
  {Thomas}, \citenamefont {Collier}, \citenamefont {Dangor}, \citenamefont
  {Divall}, \citenamefont {Foster}, \citenamefont {Gallacher}, \citenamefont
  {Hooker}, \citenamefont {Jaroszynski}, \citenamefont {Langley}, \citenamefont
  {Mori}, \citenamefont {Norreys}, \citenamefont {Tsung}, \citenamefont
  {Viskup}, \citenamefont {Walton},\ and\ \citenamefont
  {Krushelnick}}]{Mangles_Nature_2004}%
  \BibitemOpen
  \bibfield  {author} {\bibinfo {author} {\bibfnamefont {S.~P.~D.}\
  \bibnamefont {Mangles}}, \bibinfo {author} {\bibfnamefont {C.~D.}\
  \bibnamefont {Murphy}}, \bibinfo {author} {\bibfnamefont {Z.}~\bibnamefont
  {Najmudin}}, \bibinfo {author} {\bibfnamefont {A.~G.~R.}\ \bibnamefont
  {Thomas}}, \bibinfo {author} {\bibfnamefont {J.~L.}\ \bibnamefont {Collier}},
  \bibinfo {author} {\bibfnamefont {A.~E.}\ \bibnamefont {Dangor}}, \bibinfo
  {author} {\bibfnamefont {E.~J.}\ \bibnamefont {Divall}}, \bibinfo {author}
  {\bibfnamefont {P.~S.}\ \bibnamefont {Foster}}, \bibinfo {author}
  {\bibfnamefont {J.~G.}\ \bibnamefont {Gallacher}}, \bibinfo {author}
  {\bibfnamefont {C.~J.}\ \bibnamefont {Hooker}}, \bibinfo {author}
  {\bibfnamefont {D.~A.}\ \bibnamefont {Jaroszynski}}, \bibinfo {author}
  {\bibfnamefont {A.~J.}\ \bibnamefont {Langley}}, \bibinfo {author}
  {\bibfnamefont {W.~B.}\ \bibnamefont {Mori}}, \bibinfo {author}
  {\bibfnamefont {P.~A.}\ \bibnamefont {Norreys}}, \bibinfo {author}
  {\bibfnamefont {F.~S.}\ \bibnamefont {Tsung}}, \bibinfo {author}
  {\bibfnamefont {R.}~\bibnamefont {Viskup}}, \bibinfo {author} {\bibfnamefont
  {B.~R.}\ \bibnamefont {Walton}},\ and\ \bibinfo {author} {\bibfnamefont
  {K.}~\bibnamefont {Krushelnick}},\ }\bibfield  {title} {\bibinfo {title}
  {Monoenergetic beams of relativistic electrons from intense laser--plasma
  interactions},\ }\href {https://doi.org/10.1038/nature02939} {\bibfield
  {journal} {\bibinfo  {journal} {Nature}\ }\textbf {\bibinfo {volume} {431}},\
  \bibinfo {pages} {535} (\bibinfo {year} {2004})}\BibitemShut {NoStop}%
\bibitem [{\citenamefont {Gonsalves}\ \emph {et~al.}(2019)\citenamefont
  {Gonsalves}, \citenamefont {Nakamura}, \citenamefont {Daniels}, \citenamefont
  {Benedetti}, \citenamefont {Pieronek}, \citenamefont {de~Raadt},
  \citenamefont {Steinke}, \citenamefont {Bin}, \citenamefont {Bulanov},
  \citenamefont {van Tilborg}, \citenamefont {Geddes}, \citenamefont
  {Schroeder}, \citenamefont {T\'oth}, \citenamefont {Esarey}, \citenamefont
  {Swanson}, \citenamefont {Fan-Chiang}, \citenamefont {Bagdasarov},
  \citenamefont {Bobrova}, \citenamefont {Gasilov}, \citenamefont {Korn},
  \citenamefont {Sasorov},\ and\ \citenamefont
  {Leemans}}]{PhysRevLett.122.084801}%
  \BibitemOpen
  \bibfield  {author} {\bibinfo {author} {\bibfnamefont {A.~J.}\ \bibnamefont
  {Gonsalves}}, \bibinfo {author} {\bibfnamefont {K.}~\bibnamefont {Nakamura}},
  \bibinfo {author} {\bibfnamefont {J.}~\bibnamefont {Daniels}}, \bibinfo
  {author} {\bibfnamefont {C.}~\bibnamefont {Benedetti}}, \bibinfo {author}
  {\bibfnamefont {C.}~\bibnamefont {Pieronek}}, \bibinfo {author}
  {\bibfnamefont {T.~C.~H.}\ \bibnamefont {de~Raadt}}, \bibinfo {author}
  {\bibfnamefont {S.}~\bibnamefont {Steinke}}, \bibinfo {author} {\bibfnamefont
  {J.~H.}\ \bibnamefont {Bin}}, \bibinfo {author} {\bibfnamefont {S.~S.}\
  \bibnamefont {Bulanov}}, \bibinfo {author} {\bibfnamefont {J.}~\bibnamefont
  {van Tilborg}}, \bibinfo {author} {\bibfnamefont {C.~G.~R.}\ \bibnamefont
  {Geddes}}, \bibinfo {author} {\bibfnamefont {C.~B.}\ \bibnamefont
  {Schroeder}}, \bibinfo {author} {\bibfnamefont {C.}~\bibnamefont {T\'oth}},
  \bibinfo {author} {\bibfnamefont {E.}~\bibnamefont {Esarey}}, \bibinfo
  {author} {\bibfnamefont {K.}~\bibnamefont {Swanson}}, \bibinfo {author}
  {\bibfnamefont {L.}~\bibnamefont {Fan-Chiang}}, \bibinfo {author}
  {\bibfnamefont {G.}~\bibnamefont {Bagdasarov}}, \bibinfo {author}
  {\bibfnamefont {N.}~\bibnamefont {Bobrova}}, \bibinfo {author} {\bibfnamefont
  {V.}~\bibnamefont {Gasilov}}, \bibinfo {author} {\bibfnamefont
  {G.}~\bibnamefont {Korn}}, \bibinfo {author} {\bibfnamefont {P.}~\bibnamefont
  {Sasorov}},\ and\ \bibinfo {author} {\bibfnamefont {W.~P.}\ \bibnamefont
  {Leemans}},\ }\bibfield  {title} {\bibinfo {title} {Petawatt laser guiding
  and electron beam acceleration to 8 gev in a laser-heated capillary discharge
  waveguide},\ }\href {https://doi.org/10.1103/PhysRevLett.122.084801}
  {\bibfield  {journal} {\bibinfo  {journal} {Phys. Rev. Lett.}\ }\textbf
  {\bibinfo {volume} {122}},\ \bibinfo {pages} {084801} (\bibinfo {year}
  {2019})}\BibitemShut {NoStop}%
\bibitem [{\citenamefont {Aniculaesei}\ \emph {et~al.}(2023)\citenamefont
  {Aniculaesei}, \citenamefont {Ha}, \citenamefont {Yoffe}, \citenamefont
  {Labun}, \citenamefont {Milton}, \citenamefont {McCary}, \citenamefont
  {Spinks}, \citenamefont {Quevedo}, \citenamefont {Labun}, \citenamefont
  {Sain}, \citenamefont {Hannasch}, \citenamefont {Zgadzaj}, \citenamefont
  {Pagano}, \citenamefont {Franco-Altamirano}, \citenamefont {Ringuette},
  \citenamefont {Gaul}, \citenamefont {Luedtke}, \citenamefont {Tiwari},
  \citenamefont {Ersfeld}, \citenamefont {Brunetti}, \citenamefont {Ruhl},
  \citenamefont {Ditmire}, \citenamefont {Bruce}, \citenamefont {Donovan},
  \citenamefont {Downer}, \citenamefont {Jaroszynski},\ and\ \citenamefont
  {Hegelich}}]{10.1063/5.0161687}%
  \BibitemOpen
  \bibfield  {author} {\bibinfo {author} {\bibfnamefont {C.}~\bibnamefont
  {Aniculaesei}}, \bibinfo {author} {\bibfnamefont {T.}~\bibnamefont {Ha}},
  \bibinfo {author} {\bibfnamefont {S.}~\bibnamefont {Yoffe}}, \bibinfo
  {author} {\bibfnamefont {L.}~\bibnamefont {Labun}}, \bibinfo {author}
  {\bibfnamefont {S.}~\bibnamefont {Milton}}, \bibinfo {author} {\bibfnamefont
  {E.}~\bibnamefont {McCary}}, \bibinfo {author} {\bibfnamefont {M.~M.}\
  \bibnamefont {Spinks}}, \bibinfo {author} {\bibfnamefont {H.~J.}\
  \bibnamefont {Quevedo}}, \bibinfo {author} {\bibfnamefont {O.~Z.}\
  \bibnamefont {Labun}}, \bibinfo {author} {\bibfnamefont {R.}~\bibnamefont
  {Sain}}, \bibinfo {author} {\bibfnamefont {A.}~\bibnamefont {Hannasch}},
  \bibinfo {author} {\bibfnamefont {R.}~\bibnamefont {Zgadzaj}}, \bibinfo
  {author} {\bibfnamefont {I.}~\bibnamefont {Pagano}}, \bibinfo {author}
  {\bibfnamefont {J.~A.}\ \bibnamefont {Franco-Altamirano}}, \bibinfo {author}
  {\bibfnamefont {M.~L.}\ \bibnamefont {Ringuette}}, \bibinfo {author}
  {\bibfnamefont {E.}~\bibnamefont {Gaul}}, \bibinfo {author} {\bibfnamefont
  {S.~V.}\ \bibnamefont {Luedtke}}, \bibinfo {author} {\bibfnamefont
  {G.}~\bibnamefont {Tiwari}}, \bibinfo {author} {\bibfnamefont
  {B.}~\bibnamefont {Ersfeld}}, \bibinfo {author} {\bibfnamefont
  {E.}~\bibnamefont {Brunetti}}, \bibinfo {author} {\bibfnamefont
  {H.}~\bibnamefont {Ruhl}}, \bibinfo {author} {\bibfnamefont {T.}~\bibnamefont
  {Ditmire}}, \bibinfo {author} {\bibfnamefont {S.}~\bibnamefont {Bruce}},
  \bibinfo {author} {\bibfnamefont {M.~E.}\ \bibnamefont {Donovan}}, \bibinfo
  {author} {\bibfnamefont {M.~C.}\ \bibnamefont {Downer}}, \bibinfo {author}
  {\bibfnamefont {D.~A.}\ \bibnamefont {Jaroszynski}},\ and\ \bibinfo {author}
  {\bibfnamefont {B.~M.}\ \bibnamefont {Hegelich}},\ }\bibfield  {title}
  {\bibinfo {title} {{The acceleration of a high-charge electron bunch to 10
  GeV in a 10-cm nanoparticle-assisted wakefield accelerator}},\ }\href
  {https://doi.org/10.1063/5.0161687} {\bibfield  {journal} {\bibinfo
  {journal} {Matter Radiat. Extremes}\ }\textbf {\bibinfo {volume} {9}},\
  \bibinfo {pages} {014001} (\bibinfo {year} {2023})}\BibitemShut {NoStop}%
\bibitem [{\citenamefont {Poder}\ \emph {et~al.}(2018)\citenamefont {Poder},
  \citenamefont {Tamburini}, \citenamefont {Sarri}, \citenamefont {Di~Piazza},
  \citenamefont {Kuschel}, \citenamefont {Baird}, \citenamefont {Behm},
  \citenamefont {Bohlen}, \citenamefont {Cole}, \citenamefont {Corvan},
  \citenamefont {Duff}, \citenamefont {Gerstmayr}, \citenamefont {Keitel},
  \citenamefont {Krushelnick}, \citenamefont {Mangles}, \citenamefont
  {McKenna}, \citenamefont {Murphy}, \citenamefont {Najmudin}, \citenamefont
  {Ridgers}, \citenamefont {Samarin}, \citenamefont {Symes}, \citenamefont
  {Thomas}, \citenamefont {Warwick},\ and\ \citenamefont
  {Zepf}}]{Poder_PRX_2018}%
  \BibitemOpen
  \bibfield  {author} {\bibinfo {author} {\bibfnamefont {K.}~\bibnamefont
  {Poder}}, \bibinfo {author} {\bibfnamefont {M.}~\bibnamefont {Tamburini}},
  \bibinfo {author} {\bibfnamefont {G.}~\bibnamefont {Sarri}}, \bibinfo
  {author} {\bibfnamefont {A.}~\bibnamefont {Di~Piazza}}, \bibinfo {author}
  {\bibfnamefont {S.}~\bibnamefont {Kuschel}}, \bibinfo {author} {\bibfnamefont
  {C.~D.}\ \bibnamefont {Baird}}, \bibinfo {author} {\bibfnamefont
  {K.}~\bibnamefont {Behm}}, \bibinfo {author} {\bibfnamefont {S.}~\bibnamefont
  {Bohlen}}, \bibinfo {author} {\bibfnamefont {J.~M.}\ \bibnamefont {Cole}},
  \bibinfo {author} {\bibfnamefont {D.~J.}\ \bibnamefont {Corvan}}, \bibinfo
  {author} {\bibfnamefont {M.}~\bibnamefont {Duff}}, \bibinfo {author}
  {\bibfnamefont {E.}~\bibnamefont {Gerstmayr}}, \bibinfo {author}
  {\bibfnamefont {C.~H.}\ \bibnamefont {Keitel}}, \bibinfo {author}
  {\bibfnamefont {K.}~\bibnamefont {Krushelnick}}, \bibinfo {author}
  {\bibfnamefont {S.~P.~D.}\ \bibnamefont {Mangles}}, \bibinfo {author}
  {\bibfnamefont {P.}~\bibnamefont {McKenna}}, \bibinfo {author} {\bibfnamefont
  {C.~D.}\ \bibnamefont {Murphy}}, \bibinfo {author} {\bibfnamefont
  {Z.}~\bibnamefont {Najmudin}}, \bibinfo {author} {\bibfnamefont {C.~P.}\
  \bibnamefont {Ridgers}}, \bibinfo {author} {\bibfnamefont {G.~M.}\
  \bibnamefont {Samarin}}, \bibinfo {author} {\bibfnamefont {D.~R.}\
  \bibnamefont {Symes}}, \bibinfo {author} {\bibfnamefont {A.~G.~R.}\
  \bibnamefont {Thomas}}, \bibinfo {author} {\bibfnamefont {J.}~\bibnamefont
  {Warwick}},\ and\ \bibinfo {author} {\bibfnamefont {M.}~\bibnamefont
  {Zepf}},\ }\bibfield  {title} {\bibinfo {title} {Experimental signatures of
  the quantum nature of radiation reaction in the field of an ultraintense
  laser},\ }\href {https://doi.org/10.1103/PhysRevX.8.031004} {\bibfield
  {journal} {\bibinfo  {journal} {Phys. Rev. X}\ }\textbf {\bibinfo {volume}
  {8}},\ \bibinfo {pages} {031004} (\bibinfo {year} {2018})}\BibitemShut
  {NoStop}%
\bibitem [{\citenamefont {Cole}\ \emph {et~al.}(2018)\citenamefont {Cole},
  \citenamefont {Behm}, \citenamefont {Gerstmayr}, \citenamefont {Blackburn},
  \citenamefont {Wood}, \citenamefont {Baird}, \citenamefont {Duff},
  \citenamefont {Harvey}, \citenamefont {Ilderton}, \citenamefont {Joglekar},
  \citenamefont {Krushelnick}, \citenamefont {Kuschel}, \citenamefont
  {Marklund}, \citenamefont {McKenna}, \citenamefont {Murphy}, \citenamefont
  {Poder}, \citenamefont {Ridgers}, \citenamefont {Samarin}, \citenamefont
  {Sarri}, \citenamefont {Symes}, \citenamefont {Thomas}, \citenamefont
  {Warwick}, \citenamefont {Zepf}, \citenamefont {Najmudin},\ and\
  \citenamefont {Mangles}}]{Cole_PRX_2018}%
  \BibitemOpen
  \bibfield  {author} {\bibinfo {author} {\bibfnamefont {J.~M.}\ \bibnamefont
  {Cole}}, \bibinfo {author} {\bibfnamefont {K.~T.}\ \bibnamefont {Behm}},
  \bibinfo {author} {\bibfnamefont {E.}~\bibnamefont {Gerstmayr}}, \bibinfo
  {author} {\bibfnamefont {T.~G.}\ \bibnamefont {Blackburn}}, \bibinfo {author}
  {\bibfnamefont {J.~C.}\ \bibnamefont {Wood}}, \bibinfo {author}
  {\bibfnamefont {C.~D.}\ \bibnamefont {Baird}}, \bibinfo {author}
  {\bibfnamefont {M.~J.}\ \bibnamefont {Duff}}, \bibinfo {author}
  {\bibfnamefont {C.}~\bibnamefont {Harvey}}, \bibinfo {author} {\bibfnamefont
  {A.}~\bibnamefont {Ilderton}}, \bibinfo {author} {\bibfnamefont {A.~S.}\
  \bibnamefont {Joglekar}}, \bibinfo {author} {\bibfnamefont {K.}~\bibnamefont
  {Krushelnick}}, \bibinfo {author} {\bibfnamefont {S.}~\bibnamefont
  {Kuschel}}, \bibinfo {author} {\bibfnamefont {M.}~\bibnamefont {Marklund}},
  \bibinfo {author} {\bibfnamefont {P.}~\bibnamefont {McKenna}}, \bibinfo
  {author} {\bibfnamefont {C.~D.}\ \bibnamefont {Murphy}}, \bibinfo {author}
  {\bibfnamefont {K.}~\bibnamefont {Poder}}, \bibinfo {author} {\bibfnamefont
  {C.~P.}\ \bibnamefont {Ridgers}}, \bibinfo {author} {\bibfnamefont {G.~M.}\
  \bibnamefont {Samarin}}, \bibinfo {author} {\bibfnamefont {G.}~\bibnamefont
  {Sarri}}, \bibinfo {author} {\bibfnamefont {D.~R.}\ \bibnamefont {Symes}},
  \bibinfo {author} {\bibfnamefont {A.~G.~R.}\ \bibnamefont {Thomas}}, \bibinfo
  {author} {\bibfnamefont {J.}~\bibnamefont {Warwick}}, \bibinfo {author}
  {\bibfnamefont {M.}~\bibnamefont {Zepf}}, \bibinfo {author} {\bibfnamefont
  {Z.}~\bibnamefont {Najmudin}},\ and\ \bibinfo {author} {\bibfnamefont
  {S.~P.~D.}\ \bibnamefont {Mangles}},\ }\bibfield  {title} {\bibinfo {title}
  {Experimental evidence of radiation reaction in the collision of a
  high-intensity laser pulse with a laser-wakefield accelerated electron
  beam},\ }\href {https://doi.org/10.1103/PhysRevX.8.011020} {\bibfield
  {journal} {\bibinfo  {journal} {Phys. Rev. X}\ }\textbf {\bibinfo {volume}
  {8}},\ \bibinfo {pages} {011020} (\bibinfo {year} {2018})}\BibitemShut
  {NoStop}%
\bibitem [{\citenamefont {Ta~Phuoc}\ \emph {et~al.}(2012)\citenamefont
  {Ta~Phuoc}, \citenamefont {Corde}, \citenamefont {Thaury}, \citenamefont
  {Malka}, \citenamefont {Tafzi}, \citenamefont {Goddet}, \citenamefont {Shah},
  \citenamefont {Sebban},\ and\ \citenamefont {Rousse}}]{TaPhuoc2012}%
  \BibitemOpen
  \bibfield  {author} {\bibinfo {author} {\bibfnamefont {K.}~\bibnamefont
  {Ta~Phuoc}}, \bibinfo {author} {\bibfnamefont {S.}~\bibnamefont {Corde}},
  \bibinfo {author} {\bibfnamefont {C.}~\bibnamefont {Thaury}}, \bibinfo
  {author} {\bibfnamefont {V.}~\bibnamefont {Malka}}, \bibinfo {author}
  {\bibfnamefont {A.}~\bibnamefont {Tafzi}}, \bibinfo {author} {\bibfnamefont
  {J.~P.}\ \bibnamefont {Goddet}}, \bibinfo {author} {\bibfnamefont {R.~C.}\
  \bibnamefont {Shah}}, \bibinfo {author} {\bibfnamefont {S.}~\bibnamefont
  {Sebban}},\ and\ \bibinfo {author} {\bibfnamefont {A.}~\bibnamefont
  {Rousse}},\ }\bibfield  {title} {\bibinfo {title} {All-optical compton
  gamma-ray source},\ }\href {https://doi.org/10.1038/nphoton.2012.82}
  {\bibfield  {journal} {\bibinfo  {journal} {Nat. Photon.}\ }\textbf {\bibinfo
  {volume} {6}},\ \bibinfo {pages} {308–311} (\bibinfo {year}
  {2012})}\BibitemShut {NoStop}%
\bibitem [{\citenamefont {Lehe}\ \emph {et~al.}(2016)\citenamefont {Lehe},
  \citenamefont {Kirchen}, \citenamefont {Andriyash}, \citenamefont {Godfrey},\
  and\ \citenamefont {Vay}}]{Lehe2016}%
  \BibitemOpen
  \bibfield  {author} {\bibinfo {author} {\bibfnamefont {R.}~\bibnamefont
  {Lehe}}, \bibinfo {author} {\bibfnamefont {M.}~\bibnamefont {Kirchen}},
  \bibinfo {author} {\bibfnamefont {I.~A.}\ \bibnamefont {Andriyash}}, \bibinfo
  {author} {\bibfnamefont {B.~B.}\ \bibnamefont {Godfrey}},\ and\ \bibinfo
  {author} {\bibfnamefont {J.-L.}\ \bibnamefont {Vay}},\ }\bibfield  {title}
  {\bibinfo {title} {A spectral, quasi-cylindrical and dispersion-free
  particle-in-cell algorithm},\ }\href
  {https://doi.org/10.1016/j.cpc.2016.02.007} {\bibfield  {journal} {\bibinfo
  {journal} {Comp. Phys. Commun.}\ }\textbf {\bibinfo {volume} {203}},\
  \bibinfo {pages} {66–82} (\bibinfo {year} {2016})}\BibitemShut {NoStop}%
\bibitem [{\citenamefont {Diederichs}\ \emph {et~al.}(2022)\citenamefont
  {Diederichs}, \citenamefont {Benedetti}, \citenamefont {Huebl}, \citenamefont
  {Lehe}, \citenamefont {Myers}, \citenamefont {Sinn}, \citenamefont {Vay},
  \citenamefont {Zhang},\ and\ \citenamefont {Thévenet}}]{Diederichs2022}%
  \BibitemOpen
  \bibfield  {author} {\bibinfo {author} {\bibfnamefont {S.}~\bibnamefont
  {Diederichs}}, \bibinfo {author} {\bibfnamefont {C.}~\bibnamefont
  {Benedetti}}, \bibinfo {author} {\bibfnamefont {A.}~\bibnamefont {Huebl}},
  \bibinfo {author} {\bibfnamefont {R.}~\bibnamefont {Lehe}}, \bibinfo {author}
  {\bibfnamefont {A.}~\bibnamefont {Myers}}, \bibinfo {author} {\bibfnamefont
  {A.}~\bibnamefont {Sinn}}, \bibinfo {author} {\bibfnamefont {J.-L.}\
  \bibnamefont {Vay}}, \bibinfo {author} {\bibfnamefont {W.}~\bibnamefont
  {Zhang}},\ and\ \bibinfo {author} {\bibfnamefont {M.}~\bibnamefont
  {Thévenet}},\ }\bibfield  {title} {\bibinfo {title} {Hipace++: A portable,
  3d quasi-static particle-in-cell code},\ }\href
  {https://doi.org/10.1016/j.cpc.2022.108421} {\bibfield  {journal} {\bibinfo
  {journal} {Comp. Phys. Commun.}\ }\textbf {\bibinfo {volume} {278}},\
  \bibinfo {pages} {108421} (\bibinfo {year} {2022})}\BibitemShut {NoStop}%
\bibitem [{\citenamefont {Thévenet}\ \emph {et~al.}(2023)\citenamefont
  {Thévenet}, \citenamefont {Andriyash}, \citenamefont {Fedeli}, \citenamefont
  {Ferran~Pousa}, \citenamefont {Huebl}, \citenamefont {Jalas}, \citenamefont
  {Lehe},\ and\ \citenamefont {Shalloo}}]{10.5281/zenodo.10091858}%
  \BibitemOpen
  \bibfield  {author} {\bibinfo {author} {\bibfnamefont {M.}~\bibnamefont
  {Thévenet}}, \bibinfo {author} {\bibfnamefont {I.}~\bibnamefont
  {Andriyash}}, \bibinfo {author} {\bibfnamefont {L.}~\bibnamefont {Fedeli}},
  \bibinfo {author} {\bibfnamefont {A.}~\bibnamefont {Ferran~Pousa}}, \bibinfo
  {author} {\bibfnamefont {A.}~\bibnamefont {Huebl}}, \bibinfo {author}
  {\bibfnamefont {S.}~\bibnamefont {Jalas}}, \bibinfo {author} {\bibfnamefont
  {R.}~\bibnamefont {Lehe}},\ and\ \bibinfo {author} {\bibfnamefont
  {R.}~\bibnamefont {Shalloo}},\ }\href
  {https://doi.org/10.5281/ZENODO.10091858} {\bibinfo {title} {Lasy-org/lasy:
  0.4.0}} (\bibinfo {year} {2023})\BibitemShut {NoStop}%
\bibitem [{\citenamefont {Lefebvre}\ \emph {et~al.}(2003)\citenamefont
  {Lefebvre}, \citenamefont {Cochet}, \citenamefont {Fritzler}, \citenamefont
  {Malka}, \citenamefont {onard}, \citenamefont {Chemin}, \citenamefont
  {Darbon}, \citenamefont {Disdier}, \citenamefont {Faure}, \citenamefont
  {Fedotoff}, \citenamefont {Landoas}, \citenamefont {Malka}, \citenamefont
  {ot}, \citenamefont {Morel}, \citenamefont {Gloahec}, \citenamefont {Rouyer},
  \citenamefont {Rubbelynck}, \citenamefont {Tikhonchuk}, \citenamefont
  {Wrobel}, \citenamefont {Audebert},\ and\ \citenamefont
  {Rousseaux}}]{Lefebvre_NF_2003}%
  \BibitemOpen
  \bibfield  {author} {\bibinfo {author} {\bibfnamefont {E.}~\bibnamefont
  {Lefebvre}}, \bibinfo {author} {\bibfnamefont {N.}~\bibnamefont {Cochet}},
  \bibinfo {author} {\bibfnamefont {S.}~\bibnamefont {Fritzler}}, \bibinfo
  {author} {\bibfnamefont {V.}~\bibnamefont {Malka}}, \bibinfo {author}
  {\bibfnamefont {M.-M.~A.}\ \bibnamefont {onard}}, \bibinfo {author}
  {\bibfnamefont {J.-F.}\ \bibnamefont {Chemin}}, \bibinfo {author}
  {\bibfnamefont {S.}~\bibnamefont {Darbon}}, \bibinfo {author} {\bibfnamefont
  {L.}~\bibnamefont {Disdier}}, \bibinfo {author} {\bibfnamefont
  {J.}~\bibnamefont {Faure}}, \bibinfo {author} {\bibfnamefont
  {A.}~\bibnamefont {Fedotoff}}, \bibinfo {author} {\bibfnamefont
  {O.}~\bibnamefont {Landoas}}, \bibinfo {author} {\bibfnamefont
  {G.}~\bibnamefont {Malka}}, \bibinfo {author} {\bibfnamefont {V.~M.}\
  \bibnamefont {ot}}, \bibinfo {author} {\bibfnamefont {P.}~\bibnamefont
  {Morel}}, \bibinfo {author} {\bibfnamefont {M.~R.~L.}\ \bibnamefont
  {Gloahec}}, \bibinfo {author} {\bibfnamefont {A.}~\bibnamefont {Rouyer}},
  \bibinfo {author} {\bibfnamefont {C.}~\bibnamefont {Rubbelynck}}, \bibinfo
  {author} {\bibfnamefont {V.}~\bibnamefont {Tikhonchuk}}, \bibinfo {author}
  {\bibfnamefont {R.}~\bibnamefont {Wrobel}}, \bibinfo {author} {\bibfnamefont
  {P.}~\bibnamefont {Audebert}},\ and\ \bibinfo {author} {\bibfnamefont
  {C.}~\bibnamefont {Rousseaux}},\ }\bibfield  {title} {\bibinfo {title}
  {Electron and photon production from relativistic laser{\textendash}plasma
  interactions},\ }\href {https://doi.org/10.1088/0029-5515/43/7/317}
  {\bibfield  {journal} {\bibinfo  {journal} {Nucl. Fusion}\ }\textbf {\bibinfo
  {volume} {43}},\ \bibinfo {pages} {629} (\bibinfo {year} {2003})}\BibitemShut
  {NoStop}%
\bibitem [{\citenamefont {{Lobet}}\ \emph {et~al.}(2016)\citenamefont
  {{Lobet}}, \citenamefont {{d'Humi{\`e}res}}, \citenamefont {{Grech}},
  \citenamefont {{Ruyer}}, \citenamefont {{Davoine}},\ and\ \citenamefont
  {{Gremillet}}}]{Lobet_JPCS_2016}%
  \BibitemOpen
  \bibfield  {author} {\bibinfo {author} {\bibfnamefont {M.}~\bibnamefont
  {{Lobet}}}, \bibinfo {author} {\bibfnamefont {E.}~\bibnamefont
  {{d'Humi{\`e}res}}}, \bibinfo {author} {\bibfnamefont {M.}~\bibnamefont
  {{Grech}}}, \bibinfo {author} {\bibfnamefont {C.}~\bibnamefont {{Ruyer}}},
  \bibinfo {author} {\bibfnamefont {X.}~\bibnamefont {{Davoine}}},\ and\
  \bibinfo {author} {\bibfnamefont {L.}~\bibnamefont {{Gremillet}}},\
  }\bibfield  {title} {\bibinfo {title} {{Modeling of radiative and quantum
  electrodynamics effects in PIC simulations of ultra-relativistic laser-plasma
  interaction}},\ }\href {https://doi.org/10.1088/1742-6596/688/1/012058}
  {\bibfield  {journal} {\bibinfo  {journal} {J. Phys.: Conf. Ser.}\ }\textbf
  {\bibinfo {volume} {688}},\ \bibinfo {pages} {012058} (\bibinfo {year}
  {2016})}\BibitemShut {NoStop}%
\bibitem [{\citenamefont {Montefiori}\ and\ \citenamefont
  {Tamburini}(2023)}]{Montefiori2023}%
  \BibitemOpen
  \bibfield  {author} {\bibinfo {author} {\bibfnamefont {S.}~\bibnamefont
  {Montefiori}}\ and\ \bibinfo {author} {\bibfnamefont {M.}~\bibnamefont
  {Tamburini}},\ }\bibfield  {title} {\bibinfo {title} {Sfqedtoolkit: A
  high-performance library for the accurate modeling of strong-field qed
  processes in pic and monte carlo codes},\ }\href
  {https://doi.org/10.1016/j.cpc.2023.108855} {\bibfield  {journal} {\bibinfo
  {journal} {Comp. Phys. Commun.}\ }\textbf {\bibinfo {volume} {292}},\
  \bibinfo {pages} {108855} (\bibinfo {year} {2023})}\BibitemShut {NoStop}%
\bibitem [{\citenamefont {San Miguel~Claveria}\ \emph {et~al.}()\citenamefont
  {San Miguel~Claveria}, \citenamefont {Storey}, \citenamefont {Cao},
  \citenamefont {Piazza}, \citenamefont {Ekerfelt}, \citenamefont {Gessner},
  \citenamefont {Gerstmayr}, \citenamefont {Grismayer}, \citenamefont {Hogan},
  \citenamefont {Joshi}, \citenamefont {Keitel}, \citenamefont {Knetsch},
  \citenamefont {Litos}, \citenamefont {Matheron}, \citenamefont {Marsh},
  \citenamefont {Meuren}, \citenamefont {O'Shea}, \citenamefont {Reis},
  \citenamefont {Tamburini}, \citenamefont {Vranic}, \citenamefont {Wang},
  \citenamefont {Zakharova}, \citenamefont {Zhang},\ and\ \citenamefont
  {Corde}}]{claveria2023commissioningmeasurementsinitialxray}%
  \BibitemOpen
  \bibfield  {author} {\bibinfo {author} {\bibfnamefont {P.}~\bibnamefont {San
  Miguel~Claveria}}, \bibinfo {author} {\bibfnamefont {D.}~\bibnamefont
  {Storey}}, \bibinfo {author} {\bibfnamefont {G.~J.}\ \bibnamefont {Cao}},
  \bibinfo {author} {\bibfnamefont {A.~D.}\ \bibnamefont {Piazza}}, \bibinfo
  {author} {\bibfnamefont {H.}~\bibnamefont {Ekerfelt}}, \bibinfo {author}
  {\bibfnamefont {S.}~\bibnamefont {Gessner}}, \bibinfo {author} {\bibfnamefont
  {E.}~\bibnamefont {Gerstmayr}}, \bibinfo {author} {\bibfnamefont
  {T.}~\bibnamefont {Grismayer}}, \bibinfo {author} {\bibfnamefont
  {M.}~\bibnamefont {Hogan}}, \bibinfo {author} {\bibfnamefont
  {C.}~\bibnamefont {Joshi}}, \bibinfo {author} {\bibfnamefont {C.~H.}\
  \bibnamefont {Keitel}}, \bibinfo {author} {\bibfnamefont {A.}~\bibnamefont
  {Knetsch}}, \bibinfo {author} {\bibfnamefont {M.}~\bibnamefont {Litos}},
  \bibinfo {author} {\bibfnamefont {A.}~\bibnamefont {Matheron}}, \bibinfo
  {author} {\bibfnamefont {K.}~\bibnamefont {Marsh}}, \bibinfo {author}
  {\bibfnamefont {S.}~\bibnamefont {Meuren}}, \bibinfo {author} {\bibfnamefont
  {B.}~\bibnamefont {O'Shea}}, \bibinfo {author} {\bibfnamefont {D.~A.}\
  \bibnamefont {Reis}}, \bibinfo {author} {\bibfnamefont {M.}~\bibnamefont
  {Tamburini}}, \bibinfo {author} {\bibfnamefont {M.}~\bibnamefont {Vranic}},
  \bibinfo {author} {\bibfnamefont {J.}~\bibnamefont {Wang}}, \bibinfo {author}
  {\bibfnamefont {V.}~\bibnamefont {Zakharova}}, \bibinfo {author}
  {\bibfnamefont {C.}~\bibnamefont {Zhang}},\ and\ \bibinfo {author}
  {\bibfnamefont {S.}~\bibnamefont {Corde}},\ }\bibfield  {title} {\bibinfo
  {title} {Commissioning and first measurements of the initial x-ray and
  $\gamma$-ray detectors at facet-ii},\ }\href
  {https://arxiv.org/abs/2310.05535} {\ }\Eprint
  {https://arxiv.org/abs/2310.05535} {arXiv:2310.05535} \BibitemShut {NoStop}%
\bibitem [{\citenamefont {Pouyez}\ \emph {et~al.}()\citenamefont {Pouyez},
  \citenamefont {Mironov}, \citenamefont {Grismayer}, \citenamefont
  {Mercuri-Baron}, \citenamefont {Perez}, \citenamefont {Vranic}, \citenamefont
  {Riconda},\ and\ \citenamefont {Grech}}]{Pouyez2024}%
  \BibitemOpen
  \bibfield  {author} {\bibinfo {author} {\bibfnamefont {M.}~\bibnamefont
  {Pouyez}}, \bibinfo {author} {\bibfnamefont {A.~A.}\ \bibnamefont {Mironov}},
  \bibinfo {author} {\bibfnamefont {T.}~\bibnamefont {Grismayer}}, \bibinfo
  {author} {\bibfnamefont {A.}~\bibnamefont {Mercuri-Baron}}, \bibinfo {author}
  {\bibfnamefont {F.}~\bibnamefont {Perez}}, \bibinfo {author} {\bibfnamefont
  {M.}~\bibnamefont {Vranic}}, \bibinfo {author} {\bibfnamefont
  {C.}~\bibnamefont {Riconda}},\ and\ \bibinfo {author} {\bibfnamefont
  {M.}~\bibnamefont {Grech}},\ }\bibfield  {title} {\bibinfo {title}
  {Multiplicity of electron- and photon-seeded electromagnetic showers at
  multi-petawatt laser facilities},\ }\href {https://arxiv.org/abs/2402.04501}
  {\ }\Eprint {https://arxiv.org/abs/2402.04501} {arXiv:2402.04501}
  \BibitemShut {NoStop}%
\end{thebibliography}
\end{document}